\documentclass[twocolumn,footinbib,prb,superscriptaddress,reprint]{revtex4-2} %
\usepackage{xcolor}
\usepackage{amsfonts,amssymb,amsmath,mathrsfs}
\usepackage{comment}
\usepackage{graphicx}

\usepackage[colorlinks]{hyperref}%
\hypersetup{%
    plainpages=true,
    breaklinks=true,
    hypertexnames=false,
    pageanchor=true,
    colorlinks=true,
    linkcolor={blue},
    citecolor={cyan},
    urlcolor={cyan},
    anchorcolor={black}
}
\newcommand{\blk}{} %
\newcommand{\grn}{} %
\newcommand{\red}{} %
\newcommand{\blue}{} %
\newcommand{\cya}{} %

\newcommand{\xsection}[1]{\section{#1}}

\usepackage[utf8]{inputenc}

\usepackage[caption=false]{subfig}
\allowdisplaybreaks

\begin{document}

\title{\grn Gate-dependent {\red offset charge} {\blue shifts and anharmonicity} in gatemon qubits
in the weak tunneling regime}

\author{Utkan G\"{u}ng\"{o}rd\"{u}}
\email{utkan@lps.umd.edu}
\author{Rusko Ruskov}
\affiliation{Laboratory for Physical Sciences, 8050 Greenmead Drive, College Park, Maryland 20895}
\affiliation{Department of Physics, University of Maryland, College Park, Maryland 20742, USA}
\author{Silas Hoffman}
\affiliation{Laboratory for Physical Sciences, 8050 Greenmead Drive, College Park, Maryland 20895}
\author{Kyle Serniak}
\affiliation{Lincoln Laboratory, Massachusetts Institute of Technology, Lexington, MA 02421, USA}
\affiliation{Research Laboratory of Electronics, Massachusetts Institute of Technology, Cambridge, MA 02139, USA}
\author{Andrew J. Kerman}
\affiliation{Lincoln Laboratory, Massachusetts Institute of Technology, Lexington, MA 02421, USA}
\author{Charles Tahan}
\affiliation{Department of Physics, University of Maryland, College Park, Maryland 20742, USA}

\begin{abstract}

Gatemon
\grn
qubits
\blk
are based on a superconductor-quantum dot-superconductor (S-QD-S) junction which
enables in situ electrostatic tuning via a gate electrode.
For a single-channel QD this structure
\grn
gives rise to
\blk
two subgap Andreev bound states (ABSs), and generally
\grn
leads to
\blk
\cya
a richer
\blk
quantum phase dynamics as compared to conventional transmons.
In a recent work [Phys. Rev. B 111, 214503 (2025)]
we derived the quantum phase dynamics
from
\red
a many-body
\blk
treatment which leads to an effective
\blue
gate voltage-dependent
\blk
Hamiltonian
that self-consistently incorporates the phase quantization.
It predicts
(i) a renormalization of the junction's effective capacitance and
(ii) the presence of
gate voltage
\blue
and occupation-dependent
\blk
charge offsets in junctions with tunneling asymmetry.
\blue
Here,
\blk
we quantify the
observable
\red
impact
\blk
of these effects on the qubit's energy spectrum and anharmonicity,
\blue
by studying the interplay of the two Andreev branches
as a function of dot-gate voltages and junction transparencies.
We show the relation
\cya
of these predictions
\blk
to simplified gatemon models
\blk
and propose a protocol to experimentally detect the predicted charge offsets.
\end{abstract}
\maketitle

\xsection{Introduction}
Superconducting circuits incorporating hybrid
superconductor-semiconductor
Josephson junctions (JJs) have emerged as a promising platform for
the realization of compact and electrically tunable qubits
\grn
and
couplers.
\blk
A notable example is the gatemon, which shares the transmon
\grn
circuit,
\blk
a Josephson junction shunted by a large capacitance, but replaces the conventional
superconductor-insulator-superconductor (S-I-S) junction with a junction featuring
a semiconducting
region     %
typically modeled as a quantum dot (QD) that is tunnel-coupled
to the superconductor leads (S-QD-S). The presence of a gate electrode atop the semiconducting
region enables electrostatic control of    %
the quantum dot's chemical potential,
thereby allowing for in situ tuning of the qubit's energy levels.

In previous simplified models the Josephson energy matrix
associated with the two
Andreev bound states (ABSs) was introduced
phenomenologically \cite{kringhoj_anharmonicity_2018,kringhoj_suppressed_2020-1,bargerbos_observation_2020-1},
by modifying the theory of a superconducting quantum point contact
\cite{averin_coulomb_1999,zazunov_andreev_2003} (SC QPC).
For a system incorporating a S-QD-S
\red
super-semi
\blk
junction where
quantum fluctuations
\cya
on the SC phase
\blk
can be ignored
\blue
(e.g.,
\red
for
\blk
a junction shunted by a large inductance)
\blk
a closed expression for the Josephson energy
was derived
\cite{kurilovich_microwave_2021}, which reduces to the SC QPC expression in
\cya
the appropriate limit.
\blk
A contribution to the Josephson energy due to hybridization of the dot level and
the continuum states of the leads, $E_{\rm cont}$, was also derived, where
both these energies can be considered in the weak tunneling regime
($\Gamma_L,\Gamma_R \ll \Delta$) as well as in the strong tunneling regime ($\Gamma_L,\Gamma_R \gtrsim \Delta$)
where $\Gamma_L$ ($\Gamma_R$) are the
\cya
hopping energies coupling
\blk
the dot to the
\red
left (right)
\blk
lead, and
$\Delta$ is the superconducting gap in the leads.
Later on, in several studies a gatemon Hamiltonian was phenomenologically modeled based
solely on these contributions, see e.g. Refs.~\cite{fatemi_microwave_2022,fatemi_nonlinearity_2024},
\grn
in particular to explain anomalous behaviour of the gatemon anharmonicity as a function
\blk
of the QD gate voltage
~\cite{fatemi_nonlinearity_2024,Purkayashta_Frolov_2026,Sagi_Katsaros_2024,Zheng_Schonenberger_2024,
Hertel_Petersson_2022,Liu_Manucharyan_2025}.

In a recent paper, Ref.~\cite{Gungordu2025},
\grn we presented \blk
a new model for the gatemon based on
\grn a first-principles \blk
many-body treatment
of
\cya
both the super-semi junction and its circuit environment.
\blk
In this approach, an effective Hamiltonian was established using
\grn
the
\blk
path-integral formalism
\red
\cite{ambegaokar_quantum_1982,zazunov_andreev_2003},
\blk
\cya
with a
quantization of the superconducting phase derived self-consistently,
and
where the Josephson energy matrix
and continuum energy shift are in agreement with the previously
developed \cite{kurilovich_microwave_2021}.
\blk

Moreover, this model predicts new effects in the charging energy part of the Hamiltonian
that are associated with the quantum fluctuations of the SC phase.
Namely,
(i) a renormalization of the
\cya
effective capacitance across the junction,
\blk
$\delta C_{\Sigma}$,
and (ii) the emergence of two distinct charge offsets in junctions with tunneling asymmetry:
one, $\delta n_g$, that modifies the conventional
offset charge $n_g$, and
a second one, $n_z$, which depends on
the
\red
number of particles
\blk
within the junction.
Both effects are functions of the
\blue
junction gate voltage $\epsilon_g$.
\blk
We note that these
new effects
\grn have
\blk
been missing from earlier phenomenological models, see, e.g.,
Refs.~\cite{kringhoj_anharmonicity_2018,kringhoj_suppressed_2020-1,bargerbos_observation_2020-1,fatemi_nonlinearity_2024}.

In this paper, we
\grn
identify
\blk
the experimental signatures of these theoretical predictions,
focusing in particular on their influence on the gatemon
\blue
energy $n_g$-charge oscillations,
\blk
$E_{01}(n_g,\epsilon_g)$
\blue
(with
\red
its
\blk
amplitude defined as
\red
the
\blk
charge dispersion),
\blk
and
\grn
on the gatemon
\blk
anharmonicity, $\alpha(n_g,\epsilon_g)$.
We find that, within the weak tunneling regime    %
(to which this theory is restricted),
the renormalization of the capacitance has an impact on the
anharmonicity comparable to that of the energy shift
\red
$E_{\rm cont}(\phi)$
\blk
 induced by the continuum states.
Since within the weak tunneling regime the (single-channel) effective Josephson energy is of
the order of
\blue the
\blk
charging energy,
the effects
\cya
can have a significant impact on the observed
\blk
\grn
oscillations of the gatemon transition spectrum as a function of $n_g$.
\blk

An interesting {\it new effect} is
\blue
an
\blk
\blue
effective shift,
\blk
 $\delta n_{\rm eff}(\epsilon_g)$ of the
\grn
$n_g$
\blk
oscillation curves
as a function of $\epsilon_g$ in which we can recognize, in particular,
the interplay of the upper and lower
ABS energy branches as $\epsilon_g$ becomes small while the junction transparency $T(\epsilon_g)$ reaches a maximum.
We propose an experimental protocol to detect the signatures of the predicted
charge offsets by measuring the dependence of the qubit
\cya
transition
\blk
energy oscillations $E_{01}$ on $n_g$
and repeating these measurements at different values of $\epsilon_g$.

The paper is organized as follows.
\grn
In Sects.~\ref{Sec:preliminaries} and \ref{Sec:fluctuations},
\blk
we review the theoretical model of the gatemon introduced
in Ref.~\cite{Gungordu2025}.
\grn
In Sects.~\ref{Sec:transparency_and_gaps}, \ref{Sec:Josephson-energy-at-arbitrary-T}
\blk
the junction transparency, $T(\epsilon_g)$,
\blue
the
\blk
gaps between Andreev branches and
\blue
the
\blk
continuum,
and the effective Josephson energy, $E_{J,\rm eff}(T)$ for arbitrary transparency
are considered, %
as
important figures  %
for the junction.
The derivation of the boundary conditions in Sect.~\ref{Sec:Boundary_conditions},
necessary for calculating the energy spectrum,
generalizes
\grn
those
\blk
previously obtained in Ref.~\cite{vakhtel_quantum_2023},
\grn
to arbitrary tunneling rates
\blk
and for the different fermion parity sectors.
In Section \ref{Sec:Experimental_consequences}
we explore the experimental implications of the model.
\grn
In Sect.~\ref{Sec:charge-dispersion},
\blk
the charge dispersion $\delta_{01}$
[the amplitude of charge oscillations, $E_{01}(n_g)$]
is analyzed
\grn
across a range of   %
QD gate voltages $\epsilon_g$ and junction tunneling asymmetries.
\blk
In particular,
for small junction voltages and asymmetries,
$\epsilon_g, |\Gamma_{L}-\Gamma_{R}| \ll \Gamma_{L}+\Gamma_{R}$,
 we
\grn
present
\blk
 a WKB approach
(similar to Refs.~\cite{averin_coulomb_1999,vakhtel_quantum_2023})
that
\grn
provides insight
\blk
\grn
into
\blk
the interplay of
the upper and lower ABS branches.
\grn
In Sect.~\ref{Sec:charge-dispersion-shift},
\blk
\cya
we calculate
\blk
the effective  %
shift $\delta n_{\rm eff}(\epsilon_g)$ of charge oscillations that follows from
the predicted new charge offsets, $\delta n_g$ and $n_z$.
\grn
A suitable
\blk
interpretation $\delta n_{\rm eff}(\epsilon_g)$ for small and large $\epsilon_g$ is provided.
\grn
The prediction of this experimental signature is our main result
and can be used to verify the new physics of this model.
\blk
In Sect.~\ref{sec:anharmonicity_measurements},
we
examine
the effect of capacitance renormalization on the qubit anharmonicity
for moderate $\epsilon_g \sim (0.1\,\Delta, 0.7\,\Delta)$
and assess its relative significance compared to the energy shifts induced by the continuum states.
Based on the
\cya
insights
\blk
of Sect.~\ref{Sec:charge-dispersion},
we also give a qualitative explanation of the
behavior of the anharmonicity $\alpha(n_g,\epsilon_g)$ for small $\epsilon_g$.
Section \ref{Sec:Discussion} concludes the paper.

\xsection{Model}
\grn
\subsection{
\red
Overview of
\blk
predictions from path integral formalism}
\blk
\label{Sec:preliminaries}
In this subsection, we summarize the results of Ref.~\cite{Gungordu2025}.
Within the slow phase approximation \cite{ambegaokar_quantum_1982}
($\frac{e^2}{2 C_\Sigma} \ll \Delta$,
$C_\Sigma$ being the total capacitance of the dot),
and in the weak tunneling regime, $\Gamma_{L,R} \ll \sqrt{\Delta^2 - \epsilon_g^2}$,
the many-body description of the gatemon circuit
\cya
can be
\blk
reduced
\cya
to an effective Schr\"{o}dinger equation with
\blk
Hamiltonian
\blue
$\hat{H}_{\rm eff}(\hat{\phi},\hat{n}, \hat{D},\hat{D}^{\dagger})$,
with
$\hat \phi$ and $\hat n \equiv -i \partial_{\phi}$ being the conjugate operators representing
\cya
gauge-invariant phase difference across the junction
\blk
and the number of Cooper pairs transferred
\cya
through
\blk
the junction,
obeying $[\hat \phi, \hat n] = i$ \cite{Gungordu2025}.
\blk

Here, $\Gamma_{L,R}$ are the tunneling
\cya
(hopping)
\blk
rates between the dot
and the left/right lead, $\epsilon_g$ is the gate-voltage defined
\cya
energy
\blk
of the dot with respect to
the Fermi level of the leads, and $\Delta$ is the superconducting gap in the leads.
The $\hat{D}$-operator is the dot field operator in
Nambu space,
$\hat{D} = (\hat d_\uparrow, \hat d^\dagger_\downarrow)^T$
\blue
\cite{Gungordu2025},
with $\hat d^\dagger_\downarrow$ being the creation operator
of a spin-down
excitation in the dot.
\blk
It was shown in \cite{Gungordu2025} that $\hat{H}_{\rm eff}$ does not mix the even- and odd-parity sectors of the dot.
Therefore,
projecting onto the even parity occupation states, $|0\rangle$ and
$|\uparrow\downarrow\rangle = d_{\uparrow}^\dagger d_{\downarrow}^\dagger |0\rangle$
\blue
we obtain
\blk
\cite{Gungordu2025}:
\begin{align}
\label{eq:Heven}
{\hat H}_\text{even} =& \  4 {\tilde E}_C \left(-i \partial_\phi - {\tilde n}_g - n_z \eta_z  \right)^2 + \hat U_J(\phi),\\
\hat U_J(\phi) =&
\left(\frac{\Delta}{\zeta + \Gamma}\left[\Gamma \cos\frac{\phi}{2} \eta_x - \delta\Gamma \sin\frac{\phi}{2} \eta_y\right]
+ \epsilon_g \eta_z \frac{\zeta}{\zeta + \Gamma}\right)
\nonumber\\
& + E_{\rm cont}(\phi),\nonumber
\end{align}
which governs the even-parity sector
relevant for
\cya
the
\blk
gatemon.
In this expression
$\zeta \equiv \sqrt{\Delta^2 - E_A(\phi)^2}$, $\Gamma \equiv \Gamma_L + \Gamma_R$,
$\delta \Gamma \equiv \Gamma_L - \Gamma_R$,
and
$\hat U_J(\phi)$ is the Josephson energy matrix with eigenvalues defining the
lower/upper ABS energy profiles:
\begin{equation}
U_{\mp}(\phi) = \mp E_A(\phi) + E_{\rm cont}(\phi) ,
\end{equation}
with
\begin{align}
E_A(\phi) = \frac{\Delta}{\zeta + \Gamma} \sqrt{\Gamma^2 + \frac{\epsilon_g^2\zeta^2}{\Delta^2}}
\sqrt{1 - T(\epsilon_g) \sin^2\frac{\phi}{2}},
\label{eq:E_A-ABS}
\end{align}
and
\begin{equation}
T(\epsilon_g) \equiv \frac{4 \Gamma_L \Gamma_R}{ \Gamma^2 + \epsilon_g^2 \frac{\zeta^2}{\Delta^2} }
\label{eq:Transparency} .
\end{equation}
Note
that Eq.~(\ref{eq:E_A-ABS}) defines a quartic equation for $E_A(\phi)$, obtained previously
in Ref.~\cite{beenakker_superconducting_1992}.
In the derivation of
\cite{Gungordu2025} the matrix $\hat U_J(\phi) -  E_{\rm cont}(\phi)$
is non-perturbative in $\Gamma_{L,R}$
\cya
so that this model is not restricted
to the weak tunneling regime.
\blk

\cya
In this paper we focus on the weak tunneling regime,
\blk
$\Gamma_{L,R} \ll \sqrt{\Delta^2 - \epsilon_g^2}$,
\red
in which
$\zeta$ weakly depends
\blk
on $\phi$:
\begin{equation}
\zeta = \sqrt{\Delta^2-\epsilon_g^2} + {\Gamma} \frac{\epsilon_g^2}{\Delta^2 - \epsilon_g^2}
                      + {{\cal O}}\left(\frac{\Gamma_i^2}{\Delta^2 - \epsilon_g^2} \sin^2\frac{\phi}{2}  \right)
\label{eq:zeta_pert}
\end{equation}
Neglecting this $\phi$-dependence, Eq.~(\ref{eq:E_A-ABS}) can be
\cya
used as
\blk
\grn
an
\blk
approximate solution for
$E_A(\phi)$ and Eq.~(\ref{eq:Transparency}) can be interpreted as the transparency of the junction.

The hybridization of the dot level
and the continuum energies of the leads results in an energy shift, $E_{\rm cont}(\phi)$.
In the perturbative (weak tunneling) regime
\red
it is given by \cite{Gungordu2025}:
\blk
\begin{align}
\label{eq:E_Cont_pert}
E_\text{cont}(\phi) \approx & -\Gamma \frac{2}{\pi}\epsilon_g
\frac{\arcsin\frac{\epsilon_g}{\Delta}}{ \sqrt{\Delta^2 -\epsilon_g^2}} \\
& + \frac{ -2\Gamma_L \Gamma_R \Delta^2 \sin^2\frac{\phi}{2}
+ \Gamma^2 \epsilon_g^2 \left( 1+\frac{\Delta^2}{\Delta^2-\epsilon_g^2} \right)  }{ \Delta (\Delta^2 - \epsilon_g^2) }
\nonumber
\end{align}
where second order terms in $\Gamma_{L,R}$ are kept in order to capture the leading order contributions
to the supercurrent.
\red
Equation~(\ref{eq:E_Cont_pert})
\blk
is in agreement with the general non-perturbative result for
$E_{\rm cont}$ of Ref.~\cite{kurilovich_microwave_2021};
it is a good approximation for
\cya
the regime of
\blk
small or moderate $\epsilon_g$ we are interested in this paper
\footnote{
For large $\epsilon_g \gtrsim 0.5\, \Delta$
we
use
the non-perturbative version of continuum energy ~\cite{kurilovich_microwave_2021} in the form:
$E_{\text{cont}}(\phi) \equiv \mathcal E(\phi,\epsilon_g)-\mathcal E(0,0)$,
with
$\mathcal E(\phi,\epsilon_g) = \int_{-\infty}^{-\Delta}  \frac{d\epsilon}{\pi}
\arctan([\epsilon^2 - \Delta^2][\epsilon^2 - \Gamma^2 -\epsilon_g^2] - 4 \Delta^2 \Gamma_L \Gamma_R \sin^2\frac{\phi}{2},
2 \epsilon^2 \Gamma \sqrt{\epsilon^2 - \Delta^2})$,
where $\arctan(x,y)$ is the two-parameter arctangent function.
Note that the slow phase approximation leads to a restriction in the range of large
QD gate voltages:
$\epsilon_g \lesssim 0.7\, \Delta$
}.

\subsection{Junction transparency and gapped Andreev branches}
\label{Sec:transparency_and_gaps}
For the transparency
\grn
defined in Eq.~(\ref{eq:Transparency})
we note
\blk
that for $\epsilon_g \ll \Delta$, $T(\epsilon_g)$
\cya
has approximately
\blk
the Breit-Wigner form
associated with a resonance at $\epsilon_g = 0$
\grn
(this can be compared
\blk
with the phenomenology of Ref.~\cite{kringhoj_suppressed_2020-1}).
Fig.~\ref{fig:E_Jeff_largeT}, inset, shows the transparency $T(\epsilon_g)$ in the available range of $\epsilon_g$
for
\grn
a
\blk
symmetric junction, $\delta \Gamma = 0$. It reaches unity for $\epsilon_g = 0$ and decreases rapidly
for large $\epsilon_g$, according to the resonance model. In particular,
\grn
a
\blk
slower decrease of $T(\epsilon_g)$
for larger $\Gamma$ implies a wider resonance, as expected.
For
\grn
an
\blk
asymmetric junction, $\delta \Gamma \neq 0$, the transparency is scaled by a factor
$1 - \frac{\delta \Gamma^2}{\Gamma^2}$, thus reaching a maximum value less than one.

Generally, the two Andreev branches are well gapped both from each other and from the continuum states
below and above the SC gap. %
The gap to the lower/upper continuum
(at $\mp \Delta$)
is minimal at
\grn
$\phi=0$ and $2\pi$,
\blk
and is given by
$\Gamma_{A} \mp E_{\rm cont}(0)$,
where
$\Gamma_{A}(\epsilon_g)$
is the
prefactor to the last square root in Eq.~(\ref{eq:E_A-ABS}):
\begin{equation}
\Gamma_{A}(\epsilon_g) \equiv \frac{\Delta}{\zeta + \Gamma} \sqrt{\Gamma^2 + \frac{\epsilon_g^2\zeta^2}{\Delta^2}}
\equiv \sqrt{\tilde{\Gamma}^2 + \tilde{\epsilon}_g^2 }
\label{eq:Gamma_A1} ,
\end{equation}
where $\tilde{\Gamma}_{L,R} = \Gamma_{L,R} \frac{\Delta}{\zeta+\Gamma}$,
$\tilde{\epsilon}_g = \epsilon_g \frac{\zeta}{\zeta+\Gamma}$ are the re-scaled Hamiltonian parameters
(see below).
For $\epsilon_g=0$ both gaps to the continuum reach $\simeq\Gamma$.
The gap between the Andreev branches reaches
\cya
a minimum
\blk
at $\phi=\pi$
given by
$2 \Gamma_{A}(\epsilon_g)\, \sqrt{1 - T(\epsilon_g)}$.
For $\epsilon_g = 0$ it reaches a
\cya
minimum value
\blk
of
$2 \Gamma_{A}(0) \frac{\delta \Gamma}{\Gamma} \simeq 2 \delta \Gamma$,
which
\cya
goes to zero
\blk
for
\cya
a
\blk
symmetric junction.

\subsection{Effective Josephson energy at arbitrary transparency}
\label{Sec:Josephson-energy-at-arbitrary-T}
When the two ABS branches are well-gapped, which occurs in the intermediate- to low-transparency regime,
it
\grn
is
\blk
a good approximation
to consider the system ``living''
\grn
predominantly
\blk
on the lower branch
with
\cya
the phase-dependent effective potential energy:
\blk
\begin{equation}
U_{-}(\phi) = - E_A(\phi) + E_{\rm cont}(\phi)
\label{eq:lower-Andreev-branch} .
\end{equation}
\cya
This can be associated with an
\blk
effective Josephson energy, $-E_{J,{\rm eff}}\, \cos \phi$,
\cya
containing
\blk
contributions from the ABS, $E_{J,{\rm eff}}^A$, and the continuum states.
At low transparencies, the in-gap contribution can be obtained
from Eq.~\eqref{eq:E_A-ABS} as
\begin{align}
E_{J,{\rm eff}}^A \approx
\Gamma_{A}(\epsilon_g)\, \frac{T(\epsilon_g)}{4}
= \frac{\Delta}{\zeta + \Gamma}
\frac{\Gamma_L \Gamma_R}{\sqrt{\Gamma^2 + \frac{\epsilon_g^2\zeta^2}{\Delta^2}}} .
\label{eq:E_J-eff}
\end{align}

However, to elaborate on the experimental consequences of
Eq.~(\ref{eq:Heven}), one needs to extend the analysis to
large transparencies, $T \lesssim 1$, which corresponds to smaller
junction
\blue
gate
\blk
voltages, $\epsilon_g$.
A more general expression for $E_{J,{\rm eff}}^A$ for
\red
arbitrary
\blk
$T$ can be obtained
from
the first harmonic in the Fourier expansion of Eq.~\eqref{eq:E_A-ABS}
\red
(for
\blk
the lower Andreev branch;
see however, Sect.~\ref{Sec:charge-dispersion}
\red
for a discussion involving both Andreev branches):
\blk
\begin{align}
\label{eq:E_Jeff_largeT}
E_{J,{\rm eff}}^A(T) & \approx \frac{\Delta}{\zeta + \Gamma} \sqrt{\Gamma^2
+ \frac{\epsilon_g^2\zeta^2}{\Delta^2}} g_1(T),
\\
g_1(T) &= \frac{8}{3 \pi T} \left[\left(1 - \frac{T}{2} \right)E(T) - \left(1 - T \right)K(T) \right],
\nonumber
\end{align}
where $K(T)$ and $E(T)$
are
complete
elliptic integrals of the first and second kind, respectively.
Eq.~(\ref{eq:E_Jeff_largeT}) reduces to Eq.~(\ref{eq:E_J-eff}) for small transparencies
as $g_1(T){\mid_{T\ll 1}} \approx \frac{T}{4}$.

\begin{figure}
    \centering
    \includegraphics[width=0.95\columnwidth]{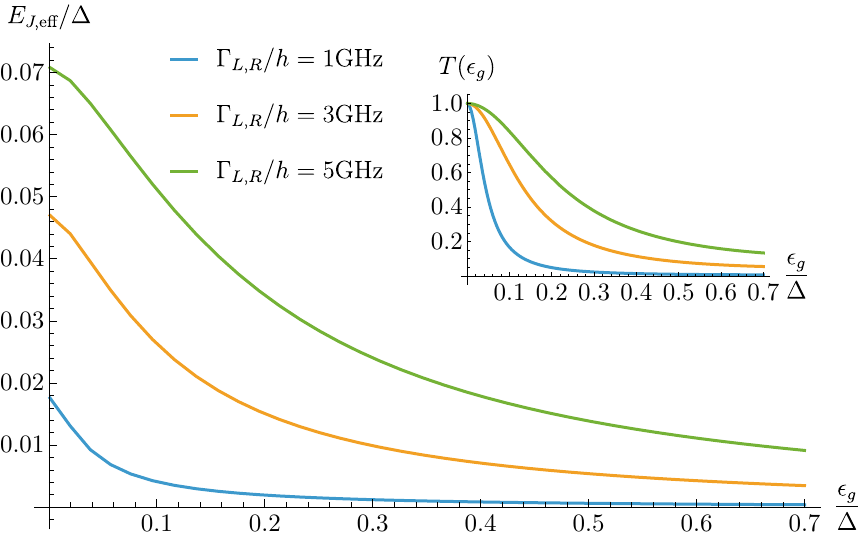}
    \caption{
        Effective Josephson energy $E_{J,{\rm eff}}(T)$,
        due to $E_A(\phi)$ and $E_{\rm cont}(\phi)$,
    and the transparency  $T(\epsilon_g)$
    as given by Eq.~(\ref{eq:Transparency}) (inset), %
            as a function of $\epsilon_g$, for $\Gamma_L = \Gamma_R$ with $T|_{\epsilon_g=0}=1$.
    For asymmetric tunneling
    \grn
    rates
    \blk
    , $\delta\Gamma \neq 0$,
    \blue
    $T(\epsilon_g)$
    \blk
     is reduced
    by the factor $1 - \frac{\delta\Gamma^2}{\Gamma^2}$.
    For parameters $\Delta/h = 45\,{\rm GHz}$, $E_C/h \approx 0.5\,{\rm GHz}$
    \cite{kringhoj_suppressed_2020-1,bargerbos_observation_2020-1},
            and $\Gamma_{L,R} = 1,3,5\, {\rm GHz}$,
    the system is not in a transmon-like regime,
    as $E_{J,{\rm eff}} \lesssim E_C$.
          }
    \label{fig:E_Jeff_largeT}
\end{figure}

Higher harmonics,
$E_{J,\text{eff},m}^A(T) \cos(m\phi)$ with $m > 1$,
of Eq.~(\ref{eq:E_A-ABS}) can be readily obtained.
These have non-negligible contributions only for large $T \approx 1$,
and rapidly decrease for smaller $T$.
The maximal strengths (at $T=1$) of the second and third harmonics are
$E_{J,{\rm eff},2}^A \approx 0.2 E_{J,{\rm eff}}^A$ and
$E_{J,{\rm eff},3}^A \approx 0.08 E_{J,{\rm eff}}^A$.
In Fig.~\ref{fig:E_Jeff_largeT},
the effective Josephson energy
(strength of the first harmonic of $E_A(\phi) - E_{\rm cont}(\phi)$,
of the lower ABS branch),
\begin{equation}
E_{J,{\rm eff}}(\epsilon_g) \approx E_{J,{\rm eff}}^A(T(\epsilon_g))
- \frac{\Gamma_L \Gamma_R \Delta}{\Delta^2 - \epsilon_g^2}  ,
\label{eq:E_J_eff}
\end{equation}
and $T(\epsilon_g)$
(inset)
are shown as a function of
$\epsilon_g \lesssim 0.7\Delta$
for
$\Gamma_{L,R}/h = 1,3,5\, {\rm GHz}$.
For a given $\Gamma_{L,R}$, both $E_{J,{\rm eff}}^A$ and $T$
attain their maxima
at $\epsilon_g = 0$ and rapidly decrease as $\epsilon_g$ increases.
The maximum value of $E_{J,{\rm eff}}^A(\epsilon_g)$ is
$\frac{4}{3\pi}\Gamma$.
For the typical experimental parameters,
$\Delta/h = 45\, {\rm GHz}$, $E_C/h \approx 0.5\, {\rm GHz}$
    \cite{kringhoj_suppressed_2020-1,bargerbos_observation_2020-1},
$E_{J,\text{eff}}(0)$ barely exceeds
or is comparable to $E_C$ when in the
\grn
single-channel
\blk
weak tunneling regime, i.e.
the gatemon system is far from being in a transmon limit
\footnote{
One can still maintain a transmon limit
by choosing much larger shunting capacitance $C_{\Sigma}$.
However, transmons with such small energy are less favorable as qubits.
\grn
Alternatively, one could consider an experimentally-relevant
multi-channel
\red
super-semi
\grn
junction,
however this is beyond the scope of the present work.
\blk
}.

\subsection{Effects of quantum fluctuations of the superconducting phase}
\label{Sec:fluctuations}
The charging energy part of $\hat{H}_{\rm even}$ in Eq.~(\ref{eq:Heven})
is
\begin{equation}
\hat{H}_C \equiv 4 \tilde E_C\left[ \hat n -  \tilde n_g(\epsilon_g) - n_z(\epsilon_g) \eta_z   \right]^2
\label{eq:E_C},
\end{equation}
and
contains several {\it new effects} \cite{Gungordu2025}
which are the main focus of this paper.
In the charging Hamiltonian,
$\tilde E_C = \frac{e^2}{2 (C_\Sigma + \delta C_\Sigma)}$ is the renormalized charging energy,
with
\begin{align}
\label{eq:deltaC}
\delta C_\Sigma(\epsilon_g) = 2e^2 \frac{\Gamma_L \Gamma_R}{\Gamma}
\frac{2 \Delta^2 + \epsilon_g^2 + 3\Delta^2 \epsilon_g
\frac{\arcsin\frac{\epsilon_g}{\Delta}}{ \sqrt{\Delta^2 -\epsilon_g^2}}   }{ \pi(\Delta^2 -\epsilon_g^2)^2}
\end{align}
being a voltage-tunable shift in  capacitance due to the continuum states.
This capacitance shift is analogous to that for
\grn
an
\blk
S-I-S based Cooper pair box \cite{ambegaokar_quantum_1982},
but can be 1-2 orders of magnitude larger for the same effective Josephson energy \cite{Gungordu2025}.

In addition, there are two new charge offsets\cite{Gungordu2025} in $\hat{H}_C$:
First, $\tilde n_g(\epsilon_g) \approx n_g + \delta n_g(\epsilon_g)$ is a renormalization of the
conventional charge offset $n_g = \frac{1}{2 e}\, \frac{C_I}{2}\, (V_L^a-V_R^a)$
(due to an applied voltage to the leads),
with
\begin{align}
\label{eq:dn_g}
\delta n_g(\epsilon_g) & \simeq
-\frac{\delta\Gamma}{2}\
\frac{\epsilon_g +  \Delta^2 \frac{ \arcsin\frac{\epsilon_g}{\Delta}}{ \sqrt{\Delta^2 -\epsilon_g^2}} }{\pi (\Delta^2 -\epsilon_g^2)}
&
\end{align}
due to the continuum states.
Second, there is an in-gap contribution to the charge offset,
\begin{align}
\label{eq:n_z}
n_z(\epsilon_g)\, \eta_z
\simeq \frac{\delta\Gamma}{4(\zeta + \Gamma)} \, \eta_z
\end{align}
which depends on the occupation of the dot, $\langle \eta_z \rangle$.
Above, the continuum contributions in $\delta C_\Sigma$ and $\delta n_g$ are perturbative results in
the leading order in $\Gamma_{L,R}$ within the weak tunneling regime, whereas
$n_z$
is an in-gap contribution,
non-perturbative in the
\cya
hopping
\blk
rates \cite{Gungordu2025}.

\subsection{Boundary conditions}
\label{Sec:Boundary_conditions}
We next derive the boundary conditions required to solve the Schr\"{o}dinger equation
\begin{align}
   \hat H_\text{even} |\psi_n(\phi) \rangle = E_n |\psi_n(\phi) \rangle,
\end{align}
for gatemon in $\phi$-space.
First, we observe that the $\propto \eta_{x,y}$ terms that cause
transitions within the even sector ($|0\rangle \leftrightarrow |\uparrow\downarrow\rangle$) do so
only by transferring a Cooper pair between the leads and the dot due to conservation of charge.
This constrains the eigenfunctions to the following form in the
\blue
$\hat n$-basis of transferred Cooper pairs
(
\cya
hereafter
we refer to these states as $|m\rangle$):
\blk
\begin{align}
|\Psi_\text{even}\rangle =
\sum_{m \in \mathbb Z \text{ or } \mathbb Z
+ \frac{1}{2} } \Psi_{m,0} |m\rangle |0\rangle
+ \Psi_{m + \frac{1}{2},\uparrow \downarrow} |m + \frac{1}{2} \rangle |\uparrow \downarrow\rangle,
\label{eq:psi-even}
\end{align}
which ensures $|0\rangle \to |\uparrow\downarrow\rangle$ is accompanied by the
change $|m \rangle \to |m + k + \frac{1}{2}\rangle$ ($k \in \mathbb Z$) by construction.
\blue
This
\blk
represents the transfer of a single Cooper pair from one of the leads onto
\blue
the dot
\blk
\cya
which results in a shift of
\blk
the eigenvalue
of $\hat n$ by $\frac{1}{2}$ since $\hat n = \frac{1}{2e}\frac{\hat Q_L - \hat Q_R}{2}$ where
\blue
$\hat Q_{L,R}$
are the charges
\blk
on each lead \cite{vakhtel_quantum_2023}.
Indeed, in $\hat H_\text{even}$,
\blue
Eq.~(\ref{eq:Heven}),
\blk
$\eta_{x,y}$ terms are always accompanied by the operators
\begin{align}
\cos\frac{\hat\phi}{2}
= \frac{1}{2}\sum_{l \in \mathbb Z \cup \mathbb Z+\frac{1}{2}} |l \rangle \langle l
+ \frac{1}{2} | + |l + \frac{1}{2}\rangle \langle l |,\nonumber\\
\sin\frac{\hat\phi}{2}
= \frac{1}{2i}\sum_{l \in \mathbb Z \cup \mathbb Z+\frac{1}{2}} |l \rangle \langle l
+ \frac{1}{2} | - |l + \frac{1}{2} \rangle \langle l |
\label{eq:cosphi2operator}
\end{align}
which exactly perform the half-integer shift in the eigenvalue of $\hat n$.
The presence of $\cos\hat\phi$-dependent terms, such as
\blue
in $\zeta(\hat\phi)$
and in $E_{\rm cont}(\hat\phi)$
\blk
\blue
(which can lead to transitions with $k \neq 0$),
\blk
is compatible with Eq.~\eqref{eq:psi-even} since such terms can only shift $m$ by an integer
\footnote{
\red
Note, that not all of the terms discussed are perturbative in $\Gamma_{L,R}$.
\blue
For the in-gap contributions to the Josephson energy and for $E_\text{cont}(\hat\phi)$
we discussed this above, in Sect.~\ref{Sec:preliminaries}.
 \red
We remark that although $\hat{H}_C$ in Ref.~\cite{Gungordu2025} was derived
\blue
in the slow phase approximation and
\red
 in the weak tunneling limit
which does not depend on phase in the leading order,
the boundary conditions are applicable to $\phi$-dependent $\hat{H}_C$
provided that $\hat{H}_C$ does not contain $\eta_{x,y}$
\blk
}.
\blue
The choice between $m \in \mathbb Z$      %
or $m \in \mathbb Z + \frac{1}{2}$ corresponds to
the parity of the leads, defined as whether the difference in
the number of
\red
quasiparticles in
\blk
the leads
is even or odd.
\blk

To obtain the boundary condition in the $\phi$-basis, we project Eq.~\eqref{eq:psi-even} onto
$|\phi\rangle$  and use $\langle \phi | l \rangle = e^{i l \phi}$
(which follows from the commutation relation $[\hat \phi,\hat n] = i$ \cite{sakurai2020modern})
and obtain
\begin{align}
& |\Psi_\text{even}(\phi)\rangle = |\phi \rangle \langle \phi|\Psi_\text{even}\rangle = \\
& \sum_{m \in \mathbb Z \text{ or } \mathbb Z
+ \frac{1}{2} } \Psi_{m,0} e^{i m \phi} |\phi \rangle |0\rangle
+ \Psi_{m + \frac{1}{2},\uparrow \downarrow} e^{i (m + \frac{1}{2})\phi}  |\phi \rangle |\uparrow \downarrow\rangle
\nonumber.
\end{align}
Returning to the basis of $\{ |\phi\rangle |\uparrow\downarrow\rangle, |\phi\rangle|0\rangle \}$,
the wavefunction is
\begin{align}
|\Psi_\text{even}(\phi)\rangle = \sum_{ m \in \mathbb Z \text{ or }  \mathbb Z + \frac{1}{2} } \begin{pmatrix}
\Psi_{m + \frac{1}{2},\uparrow \downarrow} e^{i (m + \frac{1}{2})\phi} \\
\Psi_{m,0} e^{i m \phi}
\end{pmatrix}
\end{align}
which satisfies
\begin{subequations}
\begin{eqnarray}
|\Psi_\text{even}(\phi+2\pi)\rangle &=& -\eta_z|\Psi_\text{even}(\phi)\rangle \text{ (for $m \in \mathbb Z$) }
\qquad
\label{eq:vanheck-bc}
\\
|\Psi_\text{even}(\phi+2\pi)\rangle &=& +\eta_z|\Psi_\text{even}(\phi)\rangle \text{ (for $m \in \mathbb Z
+ \frac{1}{2}$) }
\qquad
\label{eq:vanheck-bc2}
\end{eqnarray}
\end{subequations}
This result is in agreement with Ref.~\cite{vakhtel_quantum_2023} where Eq.~\eqref{eq:vanheck-bc}
was derived perturbatively in
\blue
$\Gamma_{L,R}$ in the weak $\Gamma_{L,R},\epsilon_g$ regime.
\blk

\blue
To
\blk
obtain the boundary conditions for the eigenfunctions of $\hat H_\text{odd}$
\footnote{
\red
This derivation of the boundary conditions assumes that the even- and odd-parity
sectors are well separated,
\blue
which takes place within the
perturbative in $\Gamma_{L,R}$ derivation of $\hat H_\text{odd}$
obtained in Ref.~\cite{Gungordu2025}.
This separation is further supported
\red
within the mean field treatment of
the Coulomb interaction
\blue
in the same derivation in \cite{Gungordu2025},
\red
which is valid in the deep even- or deep odd-parity regimes
~\cite{rozhkov_josephson_1999,vecino_josephson_2003}.
\blue
In general, however, the Coulomb term can mix
the even- and odd-parity sectors,
thus
necessitating an extension of the derived boundary conditions.
\blk
},
similar steps
\cya
to those
\blk
above can be retraced, by starting from the wavefunction
\begin{align}
|\Psi_\text{odd}\rangle = \sum_{m \in \mathbb Z \text{ or } \mathbb Z + \frac{1}{2} } \Psi_{m,\uparrow} |m\rangle |\uparrow\rangle + \Psi_{m , \downarrow} |m \rangle |\downarrow\rangle,
\label{eq:psi-odd}
\end{align}
which conforms to the structure of $\hat H_\text{odd}$ in which no terms are present that
can switch between
\blue
integer and half-integer $m$ (see, Ref.~\cite{Gungordu2025}).
\blk
Projecting onto the
\blue
$|\phi \rangle$
\blk
as above, we
\cya
arrive
\blk
at
\begin{align}
& |\Psi_\text{odd}(\phi)\rangle = |\phi \rangle \langle \phi|\Psi\rangle = \\
& \sum_{m \in \mathbb Z \text{ or } \mathbb Z + \frac{1}{2} } \Psi_{m,\uparrow} e^{i m \phi} |\phi \rangle |\uparrow\rangle + \Psi_{m , \downarrow} e^{i m \phi}  |\phi \rangle | \downarrow\rangle \nonumber.
\end{align}
Once again returning to the basis of $\{ |\phi\rangle |\uparrow\downarrow\rangle, |\phi\rangle|0\rangle \}$,
the wavefunction is
\begin{align}
|\Psi_\text{odd}(\phi)\rangle = \sum_{ m \in \mathbb Z \text{ or }  \mathbb Z + \frac{1}{2} } \begin{pmatrix}
\Psi_{m ,\uparrow} e^{i m \phi} \\
\Psi_{m, \downarrow} e^{i m \phi}
\end{pmatrix}
\end{align}
\begin{subequations}
\begin{align}
\label{eq:odd-bc}
|\Psi_\text{odd}(\phi+2\pi)\rangle &= +|\Psi_\text{odd}(\phi)\rangle \text{ (for $m \in \mathbb Z$) },\\
|\Psi_\text{odd}(\phi+2\pi)\rangle &= -|\Psi_\text{odd}(\phi)\rangle \text{ (for $m \in \mathbb Z + \frac{1}{2}$) } .
\end{align}
\end{subequations}

\blue
A switching between even and odd parities of the leads,
($m \in \mathbb Z$ or $m \in \mathbb Z + \frac{1}{2}$, respectively)
will occur upon a quasiparticle transfer between the leads.
Boundary conditions for the odd parity sector of the dot
junction
given in Eq.~(\ref{eq:odd-bc},b) are analogous
to the $2\pi$-periodic and $2\pi$-antiperiodic boundary conditions for the transmon wavefunctions
(see, for example, Eq.~(B2) in Ref.~\cite{catelani_relaxation_2011})  corresponding to
the even- and odd-parity sectors of the leads.
\blk

\xsection{Experimental consequences}
\label{Sec:Experimental_consequences}
In the weak tunneling regime, the effective Josephson energy
is comparable to the charging energy $E_{J,\rm eff} \lesssim E_C$
(for a single-channel device),
thus the charge dispersion of the gatemon
\grn
can
\blk
be significant.
This is a favorable regime for observing the influence of $\delta n_g$ and $n_z$ on
charge dispersion \cite{kringhoj_suppressed_2020-1} as $\epsilon_g$ or $\Gamma_{L,R}$ are varied electrically.
Voltage-dependent effects of $\delta C_\Sigma$, on the other hand,
can be observed
via the gatemon anharmonicity.

\cya
To shed light on
\blk
these experimental observables,
in this section we calculate the gatemon energy levels, $E_n$,
and eigenfunctions, $|\psi_n(\phi)\rangle$,
by numerically solving the
Schr\"{o}dinger equation with the Hamiltonian $\hat H_\text{even}$,
Eq.~\eqref{eq:Heven}, subject to the boundary condition,
Eq.~\eqref{eq:vanheck-bc2}:
\begin{equation}
|\psi_n(2\pi)\rangle = \eta_z |\psi_n(0)\rangle .
\label{eq:Boundary_Cond}
\end{equation}
\red
where
\blk
$n$ enumerates the gatemon energy levels.

\begin{figure}
    \centering
   \includegraphics[width=0.8\columnwidth]{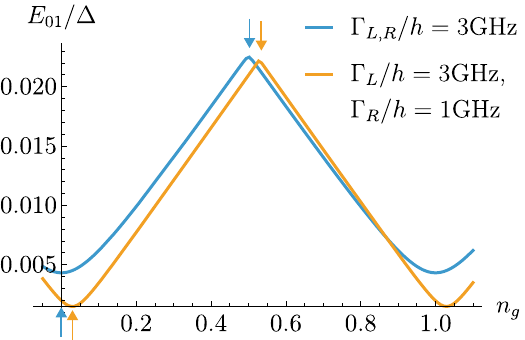}
    \caption{%
        Qubit energy oscillations,
        $E_{01}(n_g)$, obtained by solving the Schr\"{o}dinger equation
    with the Hamiltonian Eq.~(\ref{eq:Heven}) using parameters $\epsilon_g = 0.6\Delta$,
            $\Delta/h = 45\, {\rm GHz}$, $E_C/h = 0.25\, {\rm GHz}$.
    For symmetric tunneling
    \grn
    reates
    \blk
     (solid blue line), $\Gamma_i/h = 3\, {\rm GHz}$, $i=L,R$,
    the minima (maxima) of $E_{01}(n_g)$ occurs at integer (half-integer) $n_g$.
    For the asymmetric tunneling
    \grn
    rates
    \blk
     (solid
     \grn
     orange
     \blk
      line), %
    $\Gamma_L/h = 3\,{\rm GHz}$, $\Gamma_R/h = 1\,{\rm GHz}$,
            the minima and maxima are shifted due to the presence of the additional charge offsets
            $\delta n_g$ and $n_z$ from Eqs.~\eqref{eq:dn_g} and \eqref{eq:n_z},
            which is approximated by an effective charge offset, $\delta n_{\rm eff}$.
}
    \label{fig:charge-dispersion-oscillation-shift}
\end{figure}

As to the charge dispersion curves,
we calculate numerically the qubit energy splitting $E_{01}(n_g)$,
Fig.~\ref{fig:charge-dispersion-oscillation-shift},
which
oscillate as a function of the
applied voltage across the super-semi junction
($\propto n_g$, scf. Ref.~\cite{kringhoj_suppressed_2020-1}).
In Sect.~\ref{Sec:charge-dispersion}
we calculate the charge dispersion $\delta_{01}(\epsilon_g)$
(the amplitude of
oscillations of
$E_{01}(n_g) \equiv  E_{1}(n_g)- E_{0}(n_g)$ as a function of $n_g$),
for a range of available QD gate voltages, $\epsilon_g$.
In particular,
for high transparencies,
$T(\epsilon_g,\delta \Gamma) \approx 1$,
i.e. for $\epsilon_g,\delta \Gamma \ll \Gamma$
we show that charge dispersion is suppressed due to
depopulation of the lower Andreev branch,
similar to Refs.~\cite{averin_coulomb_1999,vakhtel_quantum_2023}.

We predict in Sect.~\ref{Sec:charge-dispersion-shift}
that these
charge oscillation curves will exhibit a
junction
\blue
gate
\blk
voltage
dependent shift,
$\delta n_{\rm eff}(\epsilon_g)$,
which is therefore
\grn
detectable
\blk
experimentally.
For junction parameters such that
the gap between Andreev branches is open
(transparency $T < 1$)
the shift of the gatemon
\blue
energy oscillation
\blk
curve has a simple interpretation as an additional charge
offset in an effective Cooper pair box Hamiltonian.

As for
the anharmonicity, defined from the low-lying gatemon levels, $\alpha \equiv (E_2 - E_1) - (E_1 - E_0)$,
it will be shown in Sect.~\ref{sec:anharmonicity_measurements}
that for moderate junction voltages,
$\epsilon_g \in (0.1\, \Delta, 0.7\, \Delta)$,
the effects of quantum phase fluctuations (encoded in $\delta C_\Sigma$, $\delta n_g$ and $n_z$)
will be as important as the contribution of the continuum energy shift
 $E_{\rm cont}$
(
\cya
in contract to
\blk
a simpler
model, where phase fluctuation effects are ignored \cite{fatemi_nonlinearity_2024}).
The behavior of
anharmonicity  $\alpha(\epsilon_g)$  for
small junction voltage and asymmetry is also studied and explained from
an extended WKB approach.

\subsection{Gatemon charge dispersion in the weak tunneling regime in a range of dot gate voltages}
\label{Sec:charge-dispersion}

\begin{figure*}
    \centering

    \subfloat[\label{fig:charge-dispersion}]{%
        \includegraphics[width=0.32\textwidth]{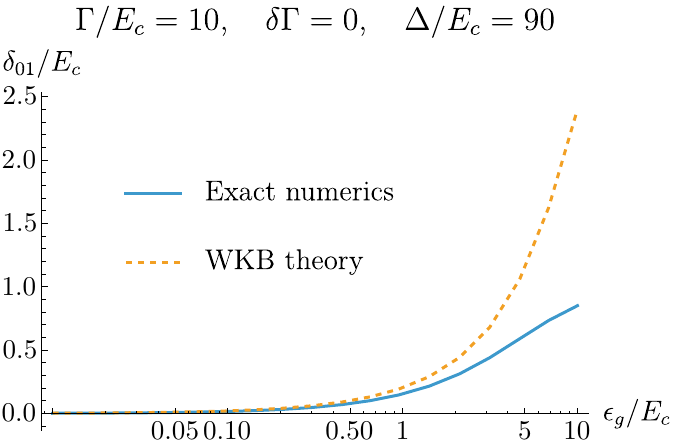}%
    }\hfill
    \subfloat[\label{fig:charge-dispersion-finite-deltaGamma}]{%
        \includegraphics[width=0.32\textwidth]{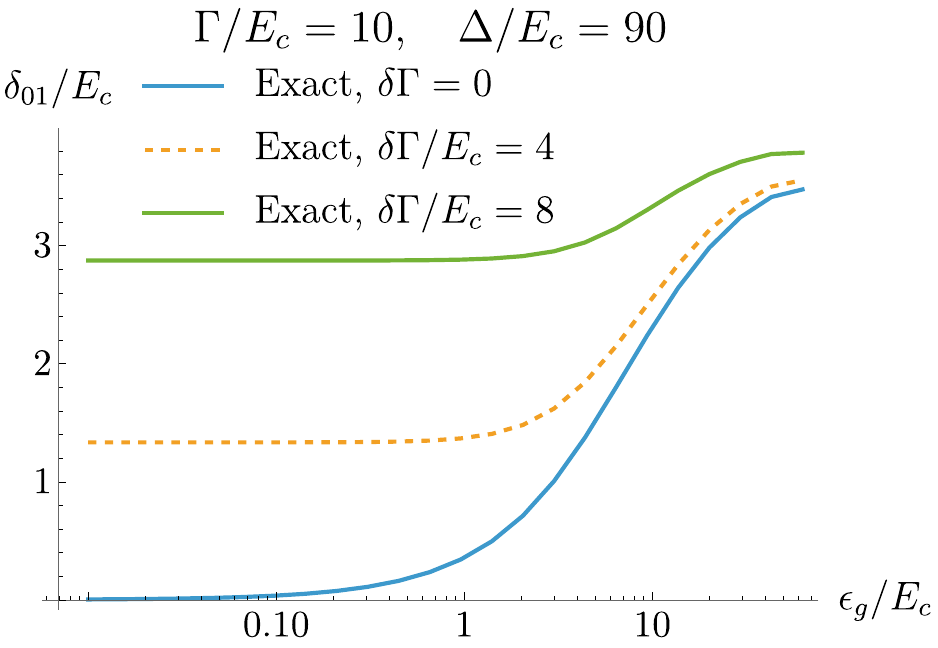}%
    }\hfill
    \subfloat[\label{fig:w-vs-epsilon_g-finite-deltaGamma}]{%
        \includegraphics[width=0.32\textwidth]{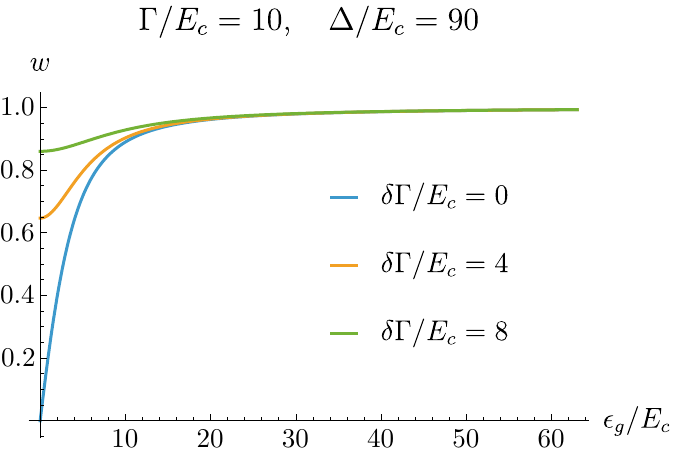}%
    }

    \caption{\blue
    Charge dispersion $\delta_{01}(\epsilon_g)$ in units of $E_C$ in (a,b),
    and
    \blue
    probability
    \blk
    amplitude $w(\epsilon_g)$ in (c),
    with $\Gamma/E_C = 10$ and $\Delta/E_C = 90$.
    (a) For symmetric tunneling
    {\grn rates}
    $\Gamma_{L,R}=5 E_C$,
    $\delta_{01}$ is calculated from the WKB theory(orange, dashed), Eq.~(\ref{eq:WKB-delta E_n}),
    and from exact numerical solutions of the Schr\"{o}dinger equation with $\hat{H}_{\rm even}$
    (blue, solid).
    (b)
    For asymmetric tunneling {\grn rates},
    $\delta\Gamma/E_C = 0,\, 4,\, 8$,
    $\delta_{01}$ is  calculated from the
    exact numerical solutions with $\hat{H}_{\rm even}$.
    Decreasing of transparency $T(\epsilon_g)$ either due to finite $\delta\Gamma$,
    or due to increasing of $\epsilon_g$
    leads to growth of $w$
    and so is the charge dispersion.
    (c)
    \blue
    Probability
    \blk
    amplitude $w$
    of remaining in the lower Andreev branch
    for $\delta\Gamma=0$
    and for finite $\delta\Gamma = 4,\,8\, E_C$. In the latter case $w$
    only slightly
    deviates from unity in the range of moderate $\epsilon_g/E_C \gtrsim 15-20$
    meaning the system is ``living'' primarily on the lower Andreev branch
    in that range.
    When $\epsilon_g, \delta\Gamma \ll \Gamma$ are both small
    $w$ go to zero and
    charge dispersion is strongly suppressed.
    \blk
    }
\end{figure*}

\cya
Figs.~\ref{fig:charge-dispersion} and \ref{fig:charge-dispersion-finite-deltaGamma}
show the calculated
\blk
charge dispersion, $\delta_{01}(\epsilon_g,\delta\Gamma,\Gamma,\Delta)$
for a range of
available QD gate voltages, $\epsilon_g$,
and junction asymmetry, $\delta\Gamma$.
Since the gap between Andreev branches closes
\grn
as the transparency approaches unity,
\blk
it
\red
is
\blk
interesting to study the effect of the upper Andreev branch
in the region of $\epsilon_g, \delta \Gamma \ll \Gamma$.
For this purpose we have adapted a WKB theory,  %
similar to
that of Refs.~\cite{averin_coulomb_1999,vakhtel_quantum_2023}.
\red
We obtain the
\blk
corrections to the gatemon energy levels,
$E_n^0 = \sqrt{8 E_C E_{J,\rm eff}}\,(n+1/2)\equiv \omega_{p,\rm eff}\,(n+1/2)$
(in a transmon-like regime),
in the form:
\begin{align}
\delta E_n(n_g)  \approx (-1)^{n+1} \frac{  w(\bar{\bar \lambda})
e^{-\tau(E_n^{0})} }{ \sigma'(E_n^{0}) } \cos(2\pi n_g + \tilde\delta)
\label{eq:WKB-delta E_n}
\end{align}
where
\grn
$\tilde\delta = {\rm arctan}\left( \frac{\Delta \delta\Gamma}{\zeta \epsilon_g}  \right)$,
\blk
and
\begin{align}
w(\bar{\bar \lambda}) & = \sqrt{\frac{2\pi}{\bar{\bar \lambda}}}\frac{e^{-\bar{\bar \lambda}}
{\bar{\bar \lambda}}^{\bar{\bar \lambda}}}{ \Gamma(\bar{\bar \lambda}) },
\label{eq:w-amplitude-lower-branch}
\end{align}
is the
\blue
probability
\blk
amplitude for the system to remain on the lower Andreev branch
when crossing the region of $\phi \approx \pi$ (under the barrier)
\cite{averin_coulomb_1999,vakhtel_quantum_2023}.
Here, $\sigma(E_n^{0})$ and $\tau(E_n^{0})$ are the WKB integrals
for the particle to be in the allowed region of the lower ABS potential around $\phi \approx 0$,
and to be in the forbidden region under the barrier, around $\phi \approx \pi$, respectively
(see Appendix \ref{App: WKB_theory} and Ref.~\cite{vakhtel_quantum_2023}).

The necessary modifications are established in the parameter
\begin{align}
\bar{\bar \lambda} & = \frac{1}{4}|r|^2
\frac{\Gamma_{A}}{\tilde{\Gamma}} \frac{\Gamma_{A}}{\bar{\bar{\Gamma}}_{A}}
\sqrt{ \frac{\bar{\bar{\Gamma}}_{A}}{E_C}},
\label{eq:lambda-parameter}
\end{align}
where
$r = \frac{\tilde{\epsilon}_{g} + i\delta \tilde{\Gamma}}{\Gamma_{A}}$ ,
$\Gamma_{A}$
is the ABS-to-continuum gap
with the re-scaled parameters, $\tilde{\Gamma}$ and $\tilde{\epsilon}_g$,
scf. Eq.~(\ref{eq:Gamma_A1}),
and the rate
\begin{equation}
\bar{\bar{\Gamma}}_{A} = \Gamma_{A} + E_{\rm cont}(\pi) - E_{\rm cont}(0) ,
\label{eq:tildeGamma_A1}
\end{equation}
is modified by the continuum energy (see Appendix \ref{App: WKB_theory}).
\red
We
\blk
note that the WKB result, Eq.~(\ref{eq:WKB-delta E_n})
requires not only transparency $T\approx 1$, but also the system to be in
\red
a transmon-like regime, $E_{J,\rm eff} \gg E_C$.
\blk
 which is not fulfilled even for small $\epsilon_g$ and $\delta\Gamma$
for a single channel device with realistic parameters.
\blue
For a single-channel device with SC gap
\red
$\Delta/h \approx 45\, {\rm GHz}$
\blk
and in the weak tunneling regime,
this can be achieved by engineering smaller $E_C$,
\blk
see Sect.\ref{Sec:Josephson-energy-at-arbitrary-T}.
Like in a transmon, charge dispersion is suppressed exponentially with the increasing ratio
\red
$E_{J,\rm eff}/E_C$.
\blk
However,
\grn
we obtain
\blk
additional suppression at $T\approx 1$ since
$w \to 0$ for $\epsilon_g, \delta \Gamma \ll \Gamma$.

In Fig.~\ref{fig:charge-dispersion}
we calculate
the amplitude of the
the charge dispersion,
defined as
$\delta_{01} = \left[{\rm max}_{n_g} E_{01}(n_g) - {\rm min}_{n_g} E_{01}(n_g)\right]/2$,
for $\delta\Gamma=0$ and $\epsilon_g$ from small to moderately large: $\epsilon_g \in (0.01 E_C, 10 E_C)$,
and compare the exact numerics from Eqs.~(\ref{eq:Heven}) and (\ref{eq:psi-even})
to the WKB result Eq.~(\ref{eq:WKB-delta E_n}).
As expected, the WKB result practically coincides with the exact one for $\epsilon_g \ll \Gamma$,
while the agreement worsens as the transparency is reduced with increasing $\epsilon_g$.
\footnote{
In general, $E_{01}(n_g)$ exhibits $1e$- and $2e$-periodic oscillations \cite{vakhtel_quantum_2023}.
However, the strength of the $1e$-periodic component becomes comparable to that of the $2e$-periodic
component only when $\epsilon_g \ll E_C$ \cite{vakhtel_quantum_2023} for which the charge dispersion
itself becomes negligible. For the parameters used in Fig.~\ref{fig:charge-dispersion},
the $1e$-periodic component is neglected.
}

In Fig.~\ref{fig:charge-dispersion-finite-deltaGamma}, we plot $\delta_{01}$
for several values of $\delta\Gamma/E_C = 0,4,8$ and for $\epsilon_g$ reaching moderate
values $\lesssim 0.7\, \Delta$.
For finite $\delta\Gamma$,
\grn
the
\blk
transparency is decreasing and charge dispersion
is finite even at $\epsilon_g \approx 0$. This opens a charge dephasing channel
for a realistic
\grn
gatemon.
\blk
For fixed $\delta\Gamma$ while increasing $\epsilon_g$, transparency
\grn
decreases
\blk
as of Eq.~(\ref{eq:Transparency}) (see
\grn
the
\blk
inset of Fig.~\ref{fig:E_Jeff_largeT}),
and
\grn
becomes
\blk
less sensitive to the factor $1 - \frac{\delta\Gamma^2}{\Gamma^2}$.
Thus, for large $\epsilon_g$ the charge dispersions for different $\delta\Gamma$
converge, Fig.~\ref{fig:charge-dispersion-finite-deltaGamma}.

In Fig.~\ref{fig:w-vs-epsilon_g-finite-deltaGamma}
we plot the
\blue
probability
\blk
amplitude $w$
of the lower Andreev branch,
illustrating that
$w$ never
\cya
reaches
\blk
zero for finite $\delta\Gamma$,
even for small $\epsilon_g$, and so is the charge dispersion
in Fig.~\ref{fig:charge-dispersion-finite-deltaGamma}.
On the other hand, for moderately large $\epsilon_g$ (low transparency)
the
\blue
probability
\blk
amplitude
$w$ saturates to unity,
Fig.~\ref{fig:w-vs-epsilon_g-finite-deltaGamma},
corresponding to the notion
of ``living on the lower Andreev branch''.
Correspondingly, the charge dispersion increases and then saturates
at some value, depending solely on the
(exponent of the) ratio of
\red
$E_{J,\rm eff}/E_C$,
\blk
which corresponds to a low-transparency
\grn
gatemon,
\blk
\blue
see inset of Fig.~\ref{fig:E_Jeff_largeT}.
\blk

In what follows, we will use the
\grn
insight
\blk
from subsection \ref{Sec:charge-dispersion},
especially the behavior of the
\blue
probability
\blk
amplitude $w$
as a function of $\epsilon_g$ and $\delta\Gamma$
to analyze the shifts of the charge oscillation curve as a function of these parameters,
which will be our main experimental prediction that follows from the
physics of Eq.~(\ref{eq:Heven}).

\subsection{
\grn
Spectroscopic signatures of
\blk
$\delta n_g$ and $n_z$}
\label{Sec:charge-dispersion-shift}
A direct observation of the new charge offsets $\delta n_g$ and $n_z$ can be made by measuring
the qubit
 energy
$E_{01}(n_g)$
(which is a 2$e$-periodic function) as $n_g$ is
varied.
For junctions with symmetric
\grn
tunneling rates
\blk
$\Gamma_L = \Gamma_R$, both $\delta n_g$ and $n_z$ vanish
and $E_{01}(n_g)$ attains its minima (maxima) at integer (half-integer) values of $n_g$,
whereas for $\Gamma_L \neq \Gamma_R$ the minima and maxima shift by an effective charge offset $\delta n_{\rm eff}$,
as shown in Fig.~\ref{fig:charge-dispersion-oscillation-shift}. %
In the medium- and low-transparency regimes
for which
ABSs are
still well gapped, an analytical description of $\delta n_{\rm eff}$ can be derived as follows.

\begin{figure*}
    \centering
    \subfloat[\label{fig:charge-dispersion-shift-vs-epsilon_g-5-1}]{%
        \includegraphics[width=0.32\textwidth]{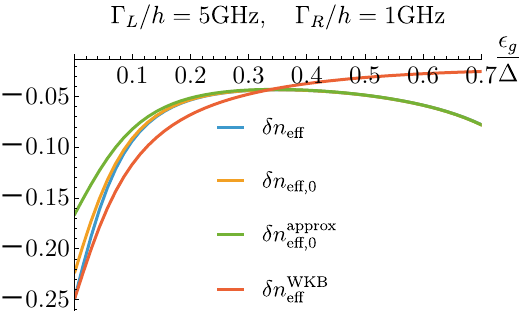}%
    }\hfill
    \subfloat[\label{fig:charge-dispersion-shift-vs-epsilon_g-6-5}]{%
        \includegraphics[width=0.32\textwidth]{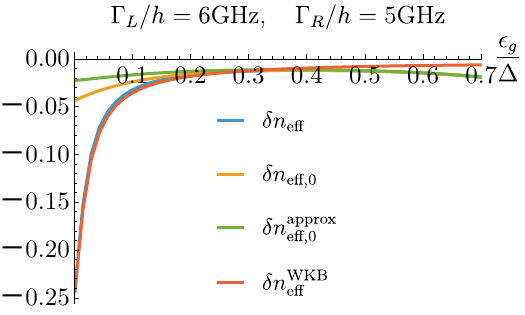}%
    }\hfill
    \subfloat[\label{fig:charge-dispersion-shift-vs-epsilon_g-6-01}]{%
        \includegraphics[width=0.32\textwidth]{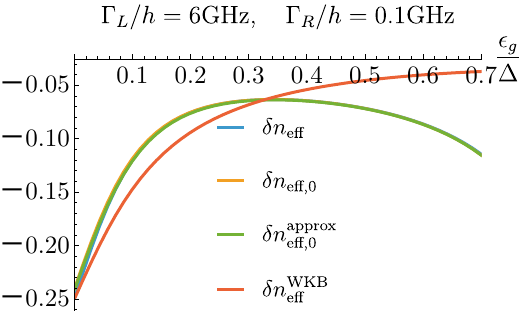}%
    }
    \caption{
    \blue
    The effective charge offset shift, $\delta n_{\rm eff}(\epsilon_g)$,
    observed in the
        qubit energy oscillations,
    Fig.~\ref{fig:charge-dispersion-oscillation-shift}, and its interpretation.
    The exact numerics $\delta n_{\rm eff}(\epsilon_g)$ (blue) is compared
    to the approximations
    $\delta n_{\rm eff,0}$, Eq.~\eqref{eq:charge_offset_effective_theory}
    {\grn (orange)}
    and
    a simplified expression $\delta n_{\rm eff,0}^{\rm approx}$,
    Eq.~\eqref{eq:charge_offset_effective_theory_approx} (green)
    that assumes the system is ``living on the lower Andreev branch'',
    expressed by the ansatz
    \red
    wavefunction,
    \blue
    Eq.~(\ref{eq:lower_branch}).
    In all three plots the WKB result $\delta n_{\rm eff}^{\rm WKB}$,
    Eq.~(\ref{eq:delta_n_WKB}), is given as the red curve:
    (a)  For
    \red
    $\delta\Gamma/h = 4\, {\rm GHz}$, $\Gamma/h = 6\, {\rm GHz}$
    \blue
    (moderate $\delta\Gamma \lesssim \Gamma$),
    the approximations practically coincide with the exact numerics
    for moderate
    $\epsilon_g \in \left(0.1\,\Delta, 0.7\, \Delta\right)$.
    These approximations worsen for small $\epsilon_g \ll \Gamma$, as the upper Andreev branch
    starts to play role, see Fig.~\ref{fig:w-vs-epsilon_g-finite-deltaGamma}.
    For $\epsilon_g \to 0$, $\delta n_{\rm eff} \to -1/4$ (or $-e/2$ in $2 e$ units)
    similar to the WKB curve (red),    %
    however the slope of the WKB curve is different, as $\delta\Gamma$ is not small.
    (b)
    For
    \red
    $\delta\Gamma/h = 1\, {\rm GHz}$, $\Gamma/h = 11\, {\rm GHz}$
    \blue
    (small $\delta\Gamma \ll \Gamma$)
    the exact numerics $\delta n_{\rm eff}(\epsilon_g)$ is well described by the
    approximations for moderate $\epsilon_g$.
    For small $\epsilon_g$ the gap between the Andreev branches closes
    as $\epsilon_g,\delta\Gamma \ll \Gamma$ ($T\approx 1$).
    Since the upper branch takes significant weight, $1-w(\epsilon_g \to 0) \to 1$,
    the $\delta n_{\rm eff}(0)$ strongly deviates from the simplified theory.
    Since   %
    WKB is applicable in this case, the exact numerics
    $\delta n_{\rm eff}(\epsilon_g)$
    is very close to the WKB result
    $\delta n_{\rm eff}^{\rm WKB}(\epsilon_g)$
    for small $\epsilon_g \lesssim 0.2\, \Delta$,
    including both the limiting value $\delta n_{\rm eff}(0) = -1/4$ and the slope.
    For higher values of $\epsilon_g$     %
    the WKB result deviates from the exact numerics, as expected.
    (c)
    For
    \red
    $\delta\Gamma/h = 5.9\, {\rm GHz}$, $\Gamma/h = 6.1\, {\rm GHz}$
    \blue
    (large $\delta\Gamma \approx \Gamma$),
    the gap between the Andreev branches is always large, and one is living on the lower branch always.
    Thus,
    the approximate theory
    for $\delta n_{\rm eff,0}$
    and
    $\delta n_{\rm eff,0}^{\rm approx}$,
    practically coincides with the exact numerics for the whole range of $\epsilon_g \in \left(0, 0.7\, \Delta \right)$.
    In all three cases, $\delta n_{\rm eff}(\epsilon_g \to 0) \to -1/4$ is a consequence of
    a low-energy theorem (for which the WKB result is only an illustration).
    For a heuristic argument see the text.
}
\end{figure*}

We first observe that in the limiting case of $\epsilon_g \gg \Gamma_{L,R}$,
the Josephson potential can be approximated as $\hat U_J \approx \epsilon_g \eta_z + E_\text{cont}(\phi)$,
\cya
which
\blk
commutes with the charge offset operator,
see Eq.~(\ref{eq:E_C}).
The lowest (qubit) energy states originate from the lower branch
of the ABSs since at low transparency the ABSs are well gapped.
Thus,
setting $\eta_z \to -1$ for the low-lying
energy eigenstates, the charging Hamiltonian, Eq.~(\ref{eq:E_C}),
becomes a
\red
scalar:
\blk
\begin{equation}
\hat{H}_{C,\rm eff} = 4 \tilde E_C(\hat n - n_g - \delta n_{\rm eff})^2
\label{eq:H_C_eff}
\end{equation}
with
\begin{equation}
\delta n_{\rm eff}(\epsilon_g)\mid_{(\epsilon_g \gg \Gamma)}\ \simeq \delta n_g(\epsilon_g) - n_z(\epsilon_g) ,
\label{eq:delta_n_eff-large-epsilon-g}
\end{equation}
which is confirmed numerically (see
Figs.~\ref{fig:charge-dispersion-shift-vs-epsilon_g-5-1},
\ref{fig:charge-dispersion-shift-vs-epsilon_g-6-5}, \ref{fig:charge-dispersion-shift-vs-epsilon_g-6-01}).

To extend this result to $\epsilon_g \lesssim \Gamma_{L,R}$,
we perform a unitary transformation $\hat U$
\grn
that
\blk
diagonalizes $\hat U_J$ as
$\hat U \hat U_J \hat U^\dagger = E_A(\phi) \sigma_z + E_\text{cont}(\phi)$
\footnote{
Explicitly,
$\hat U = \begin{pmatrix}
e^{i \Phi} \cos\frac{\Theta}{2} & \sin \frac{\Theta}{2} \\
e^{i \Phi} \sin\frac{\Theta}{2} & -\cos \frac{\Theta}{2}
\end{pmatrix}$,
$\Theta = \arccos\frac{\frac{\zeta}{\zeta + \Gamma} \epsilon_g}{E_A}$,
and
$\Phi = \arctan(\Gamma \cos\frac{\phi}{2},
-\delta\Gamma \sin\frac{\phi}{2})$.
},
where $\sigma_i$ are Pauli matrices in the ABSs basis.
In this
\cya
basis,
\blk
the Schr\"{o}dinger equation
becomes $\hat{ {\tilde H}}_\text{even} |\tilde \psi_n(\phi) \rangle = E_n |\tilde \psi_n(\phi) \rangle$,
and the rotated wavefunctions $|\tilde \psi_n(\phi) \rangle = \hat U|\psi_n(\phi) \rangle$ satisfy
the modified boundary condition
\begin{equation}
|\tilde{\psi}_n(2\pi) \rangle = -|\tilde{\psi}_n(0) \rangle .
\end{equation}
When ABSs are well gapped,
the off-diagonal terms are suppressed for low-lying qubit eigenstates, and the wavefunctions
will have a negligible component in the upper Andreev branch  %
(confirmed numerically; see also Fig.~\ref{fig:w-vs-epsilon_g-finite-deltaGamma}):
\begin{align}
| \tilde\psi_n(\phi)\rangle \approx
\red
| \tilde\psi_{n,\rm lower}(\phi)\rangle \equiv
\blk
\begin{pmatrix} 0 \\
\tilde\psi_n^\text{lower}(\phi)
\end{pmatrix}.
\label{eq:lower_branch}
\end{align}
This approximation is equivalent to replacing $\sigma_z \to -1, \sigma_{x,y} \to 0$ in the rotated Hamiltonian,
which reduces the full Hamiltonian, Eq.~(\ref{eq:Heven})
into a
\grn
scalar Hamiltonian
\blk
\cya
(equivalent to a
\blk
Cooper pair box)
with a charging energy
\red
$\hat{H}_{C,\rm eff}$,
\blk
Eq.~(\ref{eq:H_C_eff}),
and
a potential term
$\approx -E_{J,{\rm eff}}\cos\phi$, Eq.~(\ref{eq:E_Jeff_largeT}).

The additional effective charge offset can be obtained from the average of the
charge offset operator in the
\cya
transformed
\blk
charging Hamiltonian,
\begin{align}
{\hat {\tilde H}}_C
= 4 \tilde E_C\left(-i \partial_\phi -i \hat U \partial_\phi \hat U^\dagger
-\tilde n_g - n_z \hat U\eta_z \hat U^\dagger  \right)^2,
\end{align}
over the states
\red
$| \tilde\psi_{n,\rm lower}(\phi)\rangle$
\blk
($n=0,1$) as
\red
\begin{align}
\delta n_{{\rm eff},n} \equiv \delta n_g
+ \langle \tilde{\psi}_{n,\rm lower} | i \hat U \partial_\phi \hat U^{\dagger}
+ n_z \hat U \eta_z \hat U^\dagger |\tilde{\psi}_{n,\rm lower} \rangle.
\label{eq:charge_offset_effective_theory}
\end{align}
\blk
We numerically find that $\delta n_{\rm eff,0} \approx \delta n_{\rm eff,1}$,
such  that the effective charging Hamiltonian becomes a scalar, as in Eq.~(\ref{eq:H_C_eff})
with the effective charge offset $\delta n_{\rm eff,0}$ of Eq.~(\ref{eq:charge_offset_effective_theory}).
Further we note
that
the integrand of $\delta n_{\rm eff,0}$ weakly depends on $\phi$ and is concentrated around $\phi=0$,
with
negligible upper component.
These observations allow us to further simplify the expression for
the charge offset
$\delta n_{\rm eff,0}$, Eq.~(\ref{eq:charge_offset_effective_theory}) as
\begin{eqnarray}
\delta n_{\rm eff,0}^{\rm approx} \equiv \delta n_g +
\left[ \frac{\delta \Gamma}{4 \Gamma}
- \frac{\epsilon_g \zeta}{\Gamma_A(\epsilon_g) (\zeta + \Gamma)}
 \left(\frac{\delta \Gamma}{4 \Gamma} +n_z \right) \right]
 \label{eq:charge_offset_effective_theory_approx} \qquad
\end{eqnarray}
which does not require the calculation of the low-lying wavefunctions.
The accuracy of $\delta n_{\rm eff,0}^{\rm approx}$
 decreases
 with decreasing $\epsilon_g$,
although it remains a fair approximation in the
range of moderate gate voltages,
$\epsilon_g \in [0.1\Delta, 0.7\Delta]$.
as shown, e.g., in Fig.~\ref{fig:charge-dispersion-shift-vs-epsilon_g-5-1}.

In Figs.~\ref{fig:charge-dispersion-shift-vs-epsilon_g-5-1},
\ref{fig:charge-dispersion-shift-vs-epsilon_g-6-5},
\grn
and
\blk
\ref{fig:charge-dispersion-shift-vs-epsilon_g-6-01},
we compare the exact shift,
\blue
$\delta n_{\rm eff}$,
\blk
\blue
of the qubit energy oscillation
\blk
curves,
obtained as the deviation of the first minimum of $E_{01}(n_g)$ from $0$
(Fig.~\ref{fig:charge-dispersion-oscillation-shift}),
to its approximations $\delta n_{\rm eff,0}$ and $\delta n_{\rm eff,0}^{\rm approx}$,
respectively given in Eqs.~(\ref{eq:charge_offset_effective_theory})
and ~(\ref{eq:charge_offset_effective_theory_approx}).
The numerically calculated shift, %
$\delta n_{\rm eff}$ (blue),
coincides with $\delta n_{\rm eff,0}$
\blue
(orange),
\blk
$\delta n_{\rm eff,0}^{\rm approx}$ (green)
except
a narrow region of small $\epsilon_g \ll \Gamma$
where the upper Andreev branch comes into play
and the ansatz
\red
wavefunction,
\blk
Eq.~(\ref{eq:lower_branch}), is no longer valid,
scf. Fig.~\ref{fig:w-vs-epsilon_g-finite-deltaGamma}.

As expected, the approximation,
Eq.~(\ref{eq:charge_offset_effective_theory_approx}),
fails for small $\epsilon_g$, approaching the limiting value
$\delta n_{\rm eff,0}^{\rm approx}(0) = \frac{\delta \Gamma}{4 \Gamma}$,
which is negligible
for any fixed $\delta\Gamma \ll \Gamma$.
\cya
On the other hand,
\blk
the exact shift
approaches values
of
$|\delta n_{\rm eff}(\epsilon_g=0)| \simeq 0.25$
for any fixed $\delta\Gamma$.
The deviation of the exact shift from the approximations,
$\delta n_{\rm eff,0}$ (yellow), $\delta n_{\rm eff,0}^{\rm approx}$ (green)
is stronger the smaller is the gap between the Andreev branches,
$\propto \sqrt{1 - T(\epsilon_g=0)} = \frac{\delta\Gamma}{\Gamma}$;
scf. Eq.~(\ref{eq:Gamma_A1}) and
compare, Figs.~\ref{fig:charge-dispersion-shift-vs-epsilon_g-5-1},
\ref{fig:charge-dispersion-shift-vs-epsilon_g-6-5},
with
\red
$\delta\Gamma/h = \{4,1\}\,{\rm GHz}$
and
$\Gamma/h = \{6,11\}\,{\rm GHz}$,
\blk
respectively.
In the other extreme, where $\delta\Gamma \simeq \Gamma$,
see Fig.~\ref{fig:charge-dispersion-shift-vs-epsilon_g-6-01},
with
\red
$\delta\Gamma/h = 5.9\,{\rm GHz}$,
$\Gamma/h = 6.1\,{\rm GHz}$,
\blk
the two Andreev branches are well gapped for any $\epsilon_g$,
and the state is living on the lower branch, Eq.~(\ref{eq:lower-Andreev-branch}),
leading to a good accuracy of the approximations
$\delta n_{\rm eff,0}$ (yellow), $\delta n_{\rm eff,0}^{\rm approx}$ (green),
which practically coincide with the exact numerics.

For small $\epsilon_g,\delta\Gamma \ll \Gamma$ (or transparency $T \approx 1$)
the gap between the branches decreases and the participation
of the upper Andreev branch increases, see Fig.~\ref{fig:w-vs-epsilon_g-finite-deltaGamma}.
Thus, the new charge offset $n_z(\epsilon_g)\, \langle\eta_z\rangle$, Eq.~(\ref{eq:n_z}),
which is related to the
occupation of the dot, will sense this occupation.
Therefore, it will deviate from
the (scaled) WKB result (\ref{eq:WKB-delta E_n})
for the charge offset in this regime
\begin{equation}
\delta n_{\rm eff}^{\rm WKB}(\epsilon_g,\delta\Gamma) = \frac{1}{2\pi}\,
{\rm arctan}\left(\frac{\Delta\, \delta\Gamma}{\zeta\, \epsilon_g} \right)
\label{eq:delta_n_WKB} ,
\end{equation}
where the latter did not take into account the charge offsets
$\delta n_g(\epsilon_g)$, $n_z(\epsilon_g)$.
The $\delta n_{\rm eff}^{\rm WKB}(\epsilon_g,\delta\Gamma)$ is shown as the red curve
in Figs.~\ref{fig:charge-dispersion-shift-vs-epsilon_g-5-1},
\ref{fig:charge-dispersion-shift-vs-epsilon_g-6-5},
\grn
and
\blk
\ref{fig:charge-dispersion-shift-vs-epsilon_g-6-01}.
While it approaches $-0.25$ for $\epsilon_g=0$, it deviates from the exact
charge offset, $\delta n_{\rm eff}$, even for the
case of Fig.~\ref{fig:charge-dispersion-shift-vs-epsilon_g-6-5},
where the condition $\epsilon_g,\delta\Gamma \ll \Gamma$ is fulfilled
and
\grn
the
\blk
WKB approach is applicable.
Note also the different slope of $\delta n_{\rm eff}^{\rm WKB}(\epsilon_g,\delta\Gamma)$
vs. $\delta n_{\rm eff}$ for small $\epsilon_g$.

\blue
A heuristic interpretation of the numerically observed charge offset shift,
see Figs.~\ref{fig:charge-dispersion-shift-vs-epsilon_g-5-1},
\ref{fig:charge-dispersion-shift-vs-epsilon_g-6-5}, \ref{fig:charge-dispersion-shift-vs-epsilon_g-6-01},
\begin{equation}
\delta n_{\rm eff}(\epsilon_g=0) = -1/4
\end{equation}
is as follows.
\red
Consider
\blk
the gatemon
\red
wavefunction
\blue
in the form
of superposition of the states of empty dot,
$|m\rangle |0\rangle$,
 and occupied dot,
 $|m + \frac{1}{2} \rangle |\uparrow \downarrow\rangle$
 with various charge $m$,
see Eq.~(\ref{eq:psi-even}).
Here $m$ is the number of Cooper pairs transferred through the junction.
For the state with empty dot no charge is transferred, interpreted
as $\delta n_{\rm eff}=0$, and for the doubly-occupied dot (even-parity)
the net transfer of charge is 1/2, interpreted as $\delta n_{\rm eff} = -1/2$,
i.e. a charge of
\red
$-2e/2=-e$.
\blue
For $\epsilon_g = 0$ the dot's level is in resonance with the Fermi level of the
leads, which implies equal probability amplitudes for these states.
Thus, the average charge shift is
$-1/4$,
which we observed numerically.

\blk

From Figs.~\ref{fig:charge-dispersion-shift-vs-epsilon_g-5-1},
\ref{fig:charge-dispersion-shift-vs-epsilon_g-6-5}, \ref{fig:charge-dispersion-shift-vs-epsilon_g-6-01},
the additional charge offset, $\delta n_{\rm eff}(\epsilon_g)$,
can change from
\grn
a
\blk
few percent to $\lesssim 0.5$ of the electron charge in the chosen range of
junction voltages, $\epsilon_g$.
To conclude,
the qubit energy oscillations
of
$E_{01}(n_g)$ can exhibit an additional
junction
\blue
gate
\blk
voltage-dependent shift, which is therefore experimentally accessible.
This shift has a simple interpretation as an additional charge offset $\delta n_{\rm eff}(\epsilon_g)$ in
an effective Cooper pair box Hamiltonian (when the latter approximation is relevant).
\grn
It would be detectable as an $\epsilon_g$-dependent shift in the
$n_g$ periodicity of the gatemon transition spectrum.
\blk

\subsection{Detecting $\delta C_\Sigma $ in anharmonicity measurements.
\blue
Anharmonicity at small $\epsilon_g,\delta\Gamma$
\blk
}
\label{sec:anharmonicity_measurements}
The capacitance renormalization,
$\delta C_\Sigma$ has a direct impact on the qubit anharmonicity $\alpha$.
It has also been
\grn
observed
\blk
that $E_\text{cont}(\phi)$ affects $\alpha$ as well  \cite{fatemi_nonlinearity_2024}.
To quantify the significance of  $E_\text{cont}(\phi)$ and $\delta C_\Sigma$ for $\alpha$,
we calculate the anharmonicity $\alpha(\delta C_\Sigma, \delta n_g, n_z, E_\text{cont})$
by numerically solving the Schr\"{o}dinger equation in the presence of only the relevant terms
with
\grn
the
\blk
others set to zero.
Thus, we define the baseline anharmonicity as $\alpha_0 \equiv \alpha(0, 0, 0, 0)$
for which all the contributions
(static or due to quantum fluctuations of the phase)
are neglected.
In addition, we define $\alpha_{\rm stat} \equiv \alpha(0,0,0, E_{\rm cont})$,
where only static contributions are added,
and $\alpha_{\rm fluc} \equiv \alpha(\delta C_\Sigma, \delta n_g, n_z, 0)$ for which only
terms $\delta C_\Sigma, \delta n_g, n_z$ are added
(i.e., terms due to quantum phase fluctuations
that originate from the time-dependence of the phase, i.e. $\partial_\tau \phi(\tau)$).
Here, subscripts ``stat'' and ``fluc'' stand for static and
fluctuation contributions.

\begin{figure}[h]
    \centering
   \includegraphics[width=1\columnwidth]{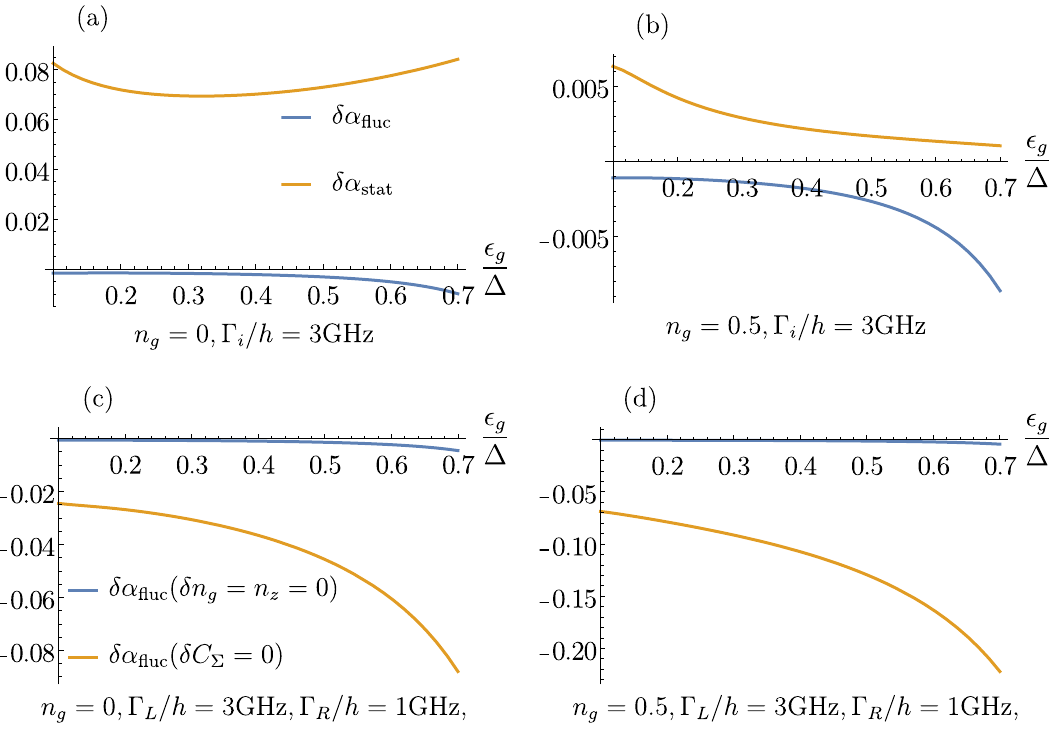}
    \caption{(a),(b) Relative anharmonicity changes $\delta \alpha_{\rm fluc}(\epsilon_g)$
    and $\delta \alpha_{\rm stat}(\epsilon_g)$ for symmetric tunnelings $\Gamma_i/h = 3\,{\rm GHz}$
    at $n_g = 0$ and $n_g = 0.5$. The effect of $\delta C_{\Sigma}$ on anharmonicity can be comparable
    to that of $E_{\rm cont}$. (c),(d) Plots for $\delta\alpha_{\rm fluc}(\delta C_\Sigma=0)$
    and $\delta\alpha_{\rm fluc}(\delta n_g=n_z=0)$ for asymmetric tunnelings $\Gamma_L/h = 3\,{\rm GHz}$,
    $\Gamma_R/h = 1\,{\rm GHz}$.
    The effects of charge offsets are found to be up to two orders of magnitude stronger.
}
    \label{fig:anharmonicity}
\end{figure}

In Fig.~\ref{fig:anharmonicity}(a),(b), we compare the relative changes in anharmonicity,
$\delta \alpha_{\rm stat} \equiv \frac{\alpha_{\rm stat} - \alpha_0}{\alpha_0}$ and
$\delta \alpha_{\rm fluc} \equiv \frac{\alpha_{\rm fluc} - \alpha_0}{\alpha_0}$, as a function of
$\epsilon_g \in [0.1\Delta, 0.7\Delta]$,
at $n_g=0$ and $n_g=0.5$,
for symmetric tunnelings (which implies $n_z = \delta n_g=0$) $\Gamma_i/h = 3\, {\rm GHz}$,
allowing us to compare the influence
 of
$\delta C_\Sigma$ and $E_\text{cont}(\phi)$ on
\grn
anharmonicity.
\blk
Despite
the fact
that the absolute values of $\delta\alpha_{\rm stat}$, $\delta\alpha_{\rm fluc}$
differ substantially for different $n_g$
(expected from the large charge dispersion),
the influence of $\delta C_\Sigma$ and $E_{\rm cont}(\phi)$
can be traced from the quantities
$\Delta\delta\alpha_{\rm stat} \equiv
|\max_{\epsilon_g} \delta\alpha_{\rm stat}(\epsilon_g) - \min_{\epsilon_g} \delta\alpha_{\rm stat}(\epsilon_g)|$
and
$\Delta\delta\alpha_{\rm fluc} \equiv
|\max_{\epsilon_g} \delta\alpha_{\rm fluc}(\epsilon_g) - \min_{\epsilon_g} \delta\alpha_{\rm fluc}(\epsilon_g)|$,
which
discard any $\epsilon_g$-independent shifts in $\alpha$.
We observe from Fig.~\ref{fig:anharmonicity}(a),(b) that for both $n_g=0$ and $n_g=0.5$,
$\Delta\delta\alpha_{\rm fluc} \sim \Delta\delta\alpha_{\rm stat}$,
showing
\grn
that
\blk
they are equally important in the interpretation of anharmonicity.
The experimentally measured behavior of $\alpha(\delta C_\Sigma, \delta n_g, n_z, E_{\rm cont})$
on $\epsilon_g$ can be used to extract
the hopping rates $\Gamma_{L,R}$
\grn
and
\blk
the capacitance renormalization $\delta C_\Sigma$.

\begin{figure*}
    \centering
    \subfloat[\label{fig:anharmonicity-high-T-ng-01}]{%
        \includegraphics[width=0.32\textwidth]{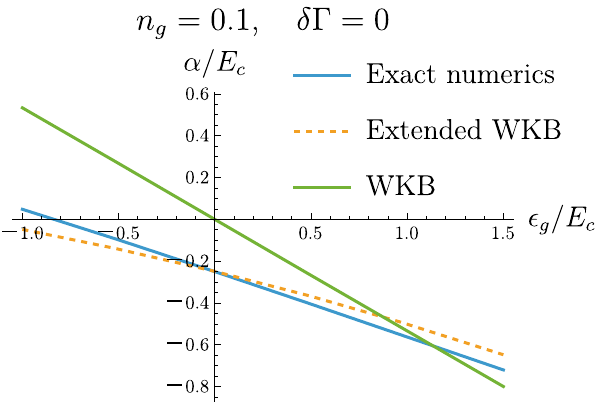}%
    }\hfill
    \subfloat[\label{fig:anharmonicity-high-T-ng-06}]{%
        \includegraphics[width=0.32\textwidth]{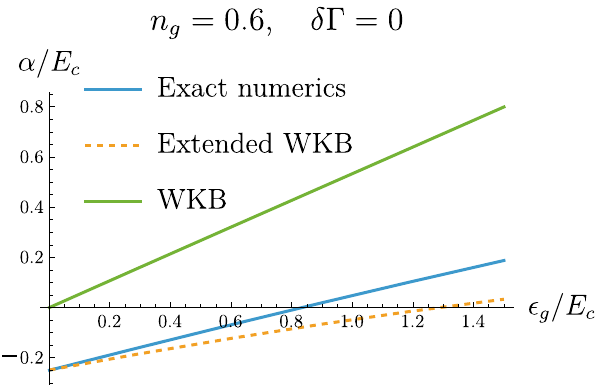}%
    }\hfill
    \subfloat[\label{fig:anharmonicity-high-T-06-delgam-01}]{%
        \includegraphics[width=0.32\textwidth]{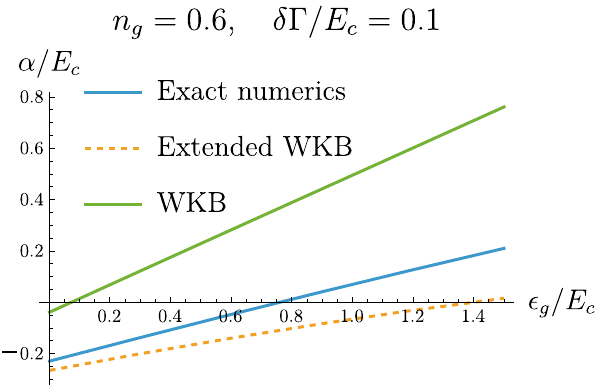}%
    }
    \caption{
    \blue
    Anharmonicity $\alpha(\epsilon_g)/E_C$ for small $\epsilon_g$,
    \blk
    $|\epsilon_g| \lesssim E_C$,
    \blue
    in a non-transmon regime, $E_C \lesssim E_{\rm J,eff}\approx 0.5 \Gamma$
    (with $\Gamma/E_C = 16$, $\Delta/E_C = 360$, $\delta\Gamma \ll E_C$).
    Exact numerics (solid blue) is compared to the WKB theory, Eq.~(\ref{eq:alpha-WKB-transmon-like}),
    (solid green), and to an extended WKB theory (dashed orange line), see Appendix~\ref{App: WKB_theory}.
    The choices for $n_g = 0.1,\,0.6$, are to avoid trivial zeros of the WKB result.
    (a)
    $\alpha(\epsilon_g)/E_C$
    for $n_g = 0.1$,
    and symmetric tunnelings $\Gamma_{L,R}  = 8\, E_C$
    (i.e., $T \approx 1$).
    The overall decrease of $|\alpha(\epsilon_g)|$ as $\epsilon_g \to 0$
    as well as the slope of $\alpha(\epsilon_g)$
    are captured by the WKB result,
    as $w(\bar{\bar \lambda}(\epsilon_g)) \to 0$, see Fig.~\ref{fig:w-vs-epsilon_g-finite-deltaGamma}
    and Eq.~(\ref{eq:alpha-WKB-transmon-like}).
    The non-zero intercept $\alpha(\epsilon_g = 0)$ is reproduced in an extended WKB approach.
    (b)
    $\alpha(\epsilon_g)/E_C$ for $n_g = 0.6$,
    and symmetric tunneling rates $\Gamma_{L,R} = 8\, E_C$
    (i.e., $T \approx 1$).
    The WKB result captures the overall decrease as $\epsilon_g \to 0$,
    as well as the slope of $\alpha(\epsilon_g)$,
    as $w(\bar{\bar \lambda}(\epsilon_g)) \to 0$.
    The change of the slope while changing from $n_g = 0.1$ to $n_g = 0.6$
    follows from the $\cos$-factor in
    Eq.~(\ref{eq:alpha-WKB-transmon-like}).
    The non-zero intercept $\alpha(\epsilon_g = 0)$ is reproduced in an extended WKB approach.
    (c)
    $\alpha(\epsilon_g)/E_C$ for $n_g = 0.6$,
    and asymmetric tunnelings: $\delta\Gamma = 0.1\, E_C$ ($\delta\Gamma \ll \Gamma$),
    corresponding to $T \approx 1$.  %
    The WKB result %
    is similar to Fig.~\ref{fig:anharmonicity-high-T-ng-06},
    however, with $w(\epsilon_g =0) \neq 0$, see Fig.~\ref{fig:w-vs-epsilon_g-finite-deltaGamma}.
    The non-zero intercept $\alpha(\epsilon_g = 0)$ is approximately reproduced in the extended WKB approach.
    \blk
}
\end{figure*}

\begin{figure}
    \centering
   \includegraphics[width=0.8\columnwidth]{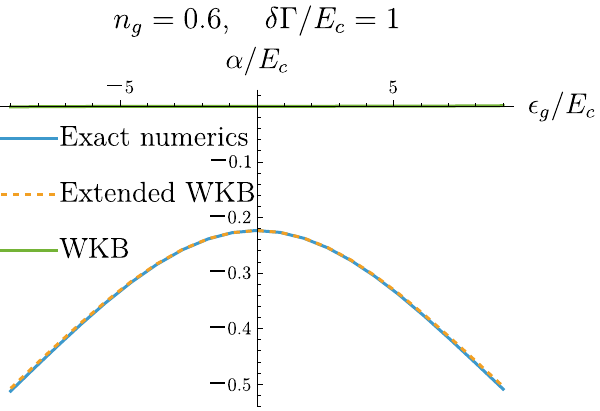}
    \caption{
    Exact numeric anharmonicity $\alpha(\epsilon_g)/E_C$
    for $n_g = 0.6$, $|\epsilon_g| \lesssim E_C$,
    and ten times smaller $E_C$ as compared to Fig.~\ref{fig:anharmonicity-high-T-06-delgam-01},
    \red
    such that
    $\Delta/E_C = 3600$,
    $\Gamma/E_C = 160$,
    \blk
    and $\delta\Gamma = E_C$,
    corresponding to $T \approx 1$ and a true transmon regime (solid blue line).
    In this case the WKB result of Eq.~(\ref{eq:alpha-WKB-transmon-like}) is exponentially
    suppressed (horizontal green line), Eq.~(\ref{eq:tau_n}), and the result is independent of $n_g$.
    An extended WKB approach (dashed orange line), Appendix~\ref{App: WKB_theory},
    practically coincides with the exact numerics, in which $\alpha(\epsilon_g) < 0$
    and it never crosses zero line (scf.
    \red
    Eq.~(10)
    \blk
    of Ref.~\cite{fatemi_nonlinearity_2024}).
    In particular, the intercept $\alpha(\epsilon_g=0)/E_C$
    \blue
    approaches
    \blk
    $-\frac{1}{4}$,
    as expected in
    \red
    the
    \blk
    short-junction limit \cite{kringhoj_anharmonicity_2018,fatemi_nonlinearity_2024}.
    \red
    See the text for further details.
    \blk
}
    \label{fig:anharmonicity-high-T-06-Ec-01}
\end{figure}

It is also worth distinguishing the quantum phase fluctuation contributions
coming from $\delta C_\Sigma$ and $\delta n_g, n_z$.
In order to isolate the charging and capacitive effects that comprise
$\delta\alpha_\text{fluc}$,
we calculate it with either $\delta C_\Sigma$ or $\delta n_g, n_z$ terms manually set to zero
in $\hat H_\text{even}$, for asymmetric tunnelings
$\Gamma_L/h = 3\, {\rm GHz}$, $\Gamma_R/h = 1\, {\rm GHz}$.
In Fig.~\ref{fig:anharmonicity}(c),(d), these two quantities,
$\delta\alpha_\text{fluc}(\delta C_\Sigma=0)$
and $\delta\alpha_\text{fluc}(\delta n_g=n_z=0)$,
are shown as a function of $\epsilon_g$,
at $n_g=0$ and $n_g=0.5$ respectively.
The effects of charge offsets
are found to be up to two orders of magnitude stronger
than the $\delta C_\Sigma$ effects,
which is expected due to the slow phase expansion \cite{Gungordu2025},
and the
strong charge dispersion in this regime.

We have so far limited our analysis on anharmonicity to medium- and low-transparency regimes,
where relative anharmonicity changes are well defined, Fig.~\ref{fig:anharmonicity}.
\grn
For
\blk
high-transparencies (and for a non-transmon-like regime of this paper),
the anharmonicity may vanish and changes sign at small $\epsilon_g$
(for the relative anharmonicity becoming ill-defined).
In the following, we study the anharmonicity $\alpha(\epsilon_g)$ plotted
for $n_g =0.1$ and $0.6$
on Figs.~\ref{fig:anharmonicity-high-T-ng-01} and \ref{fig:anharmonicity-high-T-ng-06},
for small values of junction voltage ($|\epsilon_g| \lesssim E_C$)
with $\delta \Gamma=0$,
\grn
which corresponds
\blk
to $T \approx 1$.
The exact values of $\alpha(\epsilon_g)$ (blue curves) are obtained
by numerically solving the Schr\"{o}dinger equation.
Since for this choice of parameters the WKB theory is applicable,
calculation of the anharmonicity (in a transmon-like regime) from
Eq.~(\ref{eq:WKB-delta E_n}), yields the green curves,
according to,
\begin{eqnarray}
&& \alpha_{\rm WKB}^{\rm transmon-like} =
(-1) w(\bar{\bar \lambda})
\left[\frac{ e^{-\tau(E_2^{0})} }{ \sigma'(E_2^{0})} + \frac{ 2 e^{-\tau(E_1^{0})} }{ \sigma'(E_1^{0})}
\right.  %
\nonumber\\
&&  \left.
\qquad\qquad\quad\quad { }  %
+ \frac{ e^{-\tau(E_0^{0})} }{ \sigma'(E_0^{0})}\right]
  \cos(2\pi n_g + \tilde\delta)
\label{eq:alpha-WKB-transmon-like} .
\end{eqnarray}
It shows that
$|\alpha(\epsilon_g \ll E_C)| \propto w(\bar{\bar \lambda})$ decreases to zero together with
the probability amplitude $w(\bar{\bar \lambda})$ of the lower Andreev branch,
Fig.~\ref{fig:w-vs-epsilon_g-finite-deltaGamma}.
The change
\grn
in slope
\blk
while changing from $n_g = 0.1$ to $n_g = 0.6$
is captured by the $\cos$-factor in Eq.~(\ref{eq:alpha-WKB-transmon-like}).
The latter is significant, since in a non-transmon regime
the WKB exponent
$e^{-\tau_n}$ is not suppressed by the ratio
\red
$E_{J,\rm eff}/E_C$,
\blk
see Appendix~\ref{App: WKB_theory}.

The WKB result of Eq.~(\ref{eq:alpha-WKB-transmon-like}) is not capable
\grn
of reproducing
\blk
the non-zero intercept of $\alpha(\epsilon_g=0)$.
An extended WKB approach,
which is applicable only when $T \approx 1$,
is developed in Appendix~\ref{App: WKB_theory},
where
\grn
the
\blk
anharmonicity is
given by $\alpha^{(2)}$ (see Eq.~\eqref{eq:anharmonicity-WKB}).
It coincides
with
the exact $\alpha(\epsilon_g = 0)$ at $T=1$
(dashed yellow curves in Figs.~\ref{fig:anharmonicity-high-T-ng-01} and \ref{fig:anharmonicity-high-T-ng-06}),
shows qualitative agreement when $T \approx 1$,
and approximately predicts the voltage $\epsilon_g$ at which $\alpha$ vanishes.
In Fig.~\ref{fig:anharmonicity-high-T-06-delgam-01} similar plots are shown
for $\delta \Gamma \neq 0$,
where $w(\epsilon_g =0) \neq 0$.
However,
\grn
the accuracy of Eq.~(\ref{eq:anharmonicity-WKB}) is slightly worsen
and the intercept at $\epsilon_g=0$ is overshoot.
\blk

Finally, we verified the gatemon model of Eq.~(\ref{eq:Heven})
in the deep transmon regime (by choosing
\grn
a
\blk
much smaller $E_C$),
Fig.~\ref{fig:anharmonicity-high-T-06-Ec-01}.
In this limit, the result of Eq.~(\ref{eq:alpha-WKB-transmon-like})
is exponentially suppressed (green solid line).
\blue
On the contrary,
\blk
the exact numerics and the extended WKB result,
$\alpha^{(2)} < 0$, practically coincide,
while not crossing zero
\blue
(in agreement
\blk
with
\red
Eq.~(10)
\blk
of Ref.~\cite{fatemi_nonlinearity_2024}),
and reproduce the limiting intercept of
\begin{equation}
\alpha_{\rm transmon}^{T=1}(\epsilon_g = 0)/E_C = -1/4
\label{eq:limiting intercept}
\end{equation}
expected in a short-junction limit \cite{kringhoj_anharmonicity_2018,fatemi_nonlinearity_2024}.

\blue
Indeed, in the deep transmon regime ($E_{J,\rm eff} \gg E_C$) quantum phase fluctuations are
small and the system is staying on the lower branch around
$\phi \approx 0$. Even though at $T \to 1$ the gap between the Andreev branches
is closed, transition between the branches at $\phi=\pi$ is suppressed,
since tunneling under the barrier is exponentially suppressed
as the barrier
is much higher than the system's kinetic energy.
\blk
We also checked  (not shown)
that exact numerics  at low transparency
\grn
reproduce
\blk
\blue
$\alpha_{\rm transmon}^{{\rm low}\ T}/E_C = -1$
\blk
as for the usual tunneling Hamiltonian, scf. Ref.~\cite{fatemi_nonlinearity_2024}.

\xsection{Discussion and Conclusion}
\label{Sec:Discussion}
In this work, we have investigated the experimental implications of
a gatemon model of Ref.~\cite{Gungordu2025}, derived from the underlying microscopic (many body) theory
in the weak tunneling regime for a single-channel super-semi junction.
The hybridization of the dot's level and  the continuum energies of the superconducting leads due to tunneling,
reveals itself
\grn
as
\blk
two sub-gap Andreev bound states, continuum contribution to the gatemon energy,
and
\red
$2\times 2$
\blk
matrix structure of the Josephson energy and the charging energy,
derived self-consistently \cite{Gungordu2025}.
The effects of quantum fluctuations of the superconducting phase result in capacitance
renormalization, $\delta C_{\Sigma}(\epsilon_g,\Gamma_{L,R})$, of the gatemon circuit,
and to new charge offset renormalizations, $\delta n_g(\epsilon_g,\Gamma_{L,R})$ and
$n_z(\epsilon_g,\Gamma_{L,R})$,
all depending on the QD voltage $\epsilon_g$ and the
\blue
hopping
\blk
rates to the dot, $\Gamma_{L,R}$.
Experimental control
\grn
of
\blk
$\epsilon_g$
\grn
as
\blk
in the experiments of Refs.~\cite{kringhoj_suppressed_2020-1,bargerbos_observation_2020-1},
and on $\Gamma_{L,R}$
\grn
as
\blk
in the experiment of
Ref.~\cite{Chatterjee-natmaterials2025}
could be the key
\grn
to
\blk
the experimental observation of these new quantum fluctuation effects
\grn
arising from quantum fluctuations of the phase.
\blk

In this paper, we
\grn
focus
\blk
on the gatemon
\blue
energy
\blk
oscillations (charge oscillations),
\blue
$E_{01}(n_g)$,
\blk
as a function of the applied voltage
\grn
across
\blk
the super-semi junction, $n_g \propto V_L^a - V_R^a$.
We solve the Schr\"{o}dinger equation for the Hamiltonian, $\hat{H}_{\rm even}$,
with boundary conditions derived for the even parity sector,
not assuming small
\blue
hopping
\blk
rates,
see Sect.~\ref{Sec:Boundary_conditions}.
Our main observation is that for
\grn
an
\blk
asymmetric junction with $\delta\Gamma \neq 0$
the new charge offsets, $\delta n_g$
\grn
and
\blk
 $n_z$, induce
an observable charge shift in the oscillation curve, $\delta n_{\rm eff}$.
An experiment similar to that of Ref.~\cite{kringhoj_suppressed_2020-1}
can be performed in the asymmetric weak tunneling regime
to observe the shift of the peaks of
\grn
the
\blk
qubit energy oscillations $E_{01}(n_g)$
as a function of the junction voltage $\epsilon_g$.

In a range of parameters where the two Andreev branches are well gapped
$\delta n_{\rm eff}$
\red
can be interpreted as the
\blk
charge offset in an
\red
effective
\blk
scalar Cooper pair box Hamiltonian, see Sect.~\ref{Sec:charge-dispersion-shift}.
This corresponds to a simple model of primarily ``living on the lower Andreev branch''.
In the range of small $\epsilon_g, \delta\Gamma \ll \Gamma$, where the junction transparency
$T \approx 1$ and the gap between the branches
\red
closes,
\blk
the effects of the upper Andreev branch
are coming into play.
This is elucidated in a WKB theory for the charge dispersion, $\delta_{01}$
(the amplitude of the charge oscillation curve)
adapted from Refs.~\cite{averin_coulomb_1999,vakhtel_quantum_2023}
via suitable
\red
rescaling.
\blk

The intuition behind the
\grn
simple model above
\blk
is supported by the
key quantity of the WKB theory \cite{averin_coulomb_1999,vakhtel_quantum_2023}
-- the probability amplitude $w(\epsilon_g)$ of remaining on the lower branch.
The latter was associated with a Landau-Zener transition between the branches
due to a movement under the
\red
potential
\blk
barrier
\blue
near the anti-crossing
\blk
at $\phi = \pi$ \cite{averin_coulomb_1999}.
\grn
The exact
\blk
numerical calculations of $\delta_{01}$ for asymmetric junction reveal
non-vanishing charge dispersion on
\grn
a
\blk
par
with the behavior of $w(\epsilon_g)$,
see Sect.~\ref{Sec:charge-dispersion}.
Thus, an asymmetric junction is generally more prone to charge dephasing noise,
especially when the gatemon is not in a true
transmon limit.

Our analysis shows that the contribution of capacitance renormalization, $\delta C_\Sigma(\epsilon_g)$,
to the qubit anharmonicity $\alpha$ can be comparable to that
\grn
of
\blk
the continuum energy term,
$E_\text{cont}(\phi)$,
 which was proposed as an explanation
\grn
for
\blk
recently observed anomalies in
gatemon anharmonicities \cite{fatemi_nonlinearity_2024, kringhoj_suppressed_2020-1, fatemi_microwave_2022}.
This result underscores the need to include $\delta C(\epsilon_g)$ when interpreting anharmonicity measurements
and designing high-fidelity gate operations.
Additionally, we studied the behavior of the gatemon
\red
anharmonicity
\blk
in the transition
from a non-transmon-like regime
to a true transmon regime in the region of large transparencies, $T \approx 1$.
An extended WKB analysis allows
\grn
us
\blk
to predict and understand the non-zero intercept, $\alpha(\epsilon_g=0)$
in these regimes;
see Sect.~\ref{sec:anharmonicity_measurements}.

Due to the perturbative nature of the underlying model in Ref.~\cite{Gungordu2025}, our predictions are limited to
the weak tunneling regime. Extending these results to the regime of intermediate and strong tunnel couplings will be
important to understand a wider range of devices.

\xsection{Acknowledgements}
\grn
We thank T. Hazard, K. Sardashti, and C. J. K. Richardson for valuable discussions.
\blk
This research was funded by the LPS Qubit Collaboratory
and in part under Air Force Contract No. FA8702-15-D-0001.
Any opinions, findings, conclusions,
 or recommendations expressed in this material are those of the authors and do not
necessarily reflect the views of the US Air Force or the US Government.

\appendix

\xsection{WKB Theory}
\label{App: WKB_theory}

\subsection{WKB integrals}
In this section, we consider the WKB theory
of $\hat{H}_{\rm even}$, Eq.~(\ref{eq:Heven}), in which the Josephson matrix potential, $U_J(\phi)$,
establishes two Andreev branches, $U_{\mp} = \mp E_A(\phi) + E_{\rm cont}(\phi)$.
One is interested in a regime of near perfect transparency ($T \approx 1$)
where matching conditions between the different WKB solutions around the region
of minimal gap (at $\phi=\pi$) between the branches $U_{\mp}$ can be applied,
similar to Refs.~\cite{averin_coulomb_1999,vakhtel_quantum_2023}.
We note
that the gatemon system analyzed in Ref.~\cite{vakhtel_quantum_2023}
formally corresponds to the $\Delta \to \infty$ limit of $\hat H_{\rm even}$.
For finite $\Delta$
the matrix potential $\hat{U}_J(\phi) - E_{\rm cont}(\phi)$
\red
can be related
\blk
to that of Ref.~\cite{vakhtel_quantum_2023}
\red
through the following rescalings:
\blk
\begin{align}
\left(\tilde{\Gamma},\, \delta\tilde{\Gamma} \right) &
= \frac{\Delta}{\zeta+\Gamma} \left(\Gamma,\, \delta\Gamma \right) \ \ \
\tilde{\epsilon}_{g} = \frac{\zeta}{\zeta + \Gamma} \epsilon_g
\label{eq:rescalings} .
\end{align}
\blue
It will be convenient to modify the lower ABS branch, $U_{-}(\phi)$,
by its minimal gap to
continuum, scf. Eq.~(\ref{eq:Gamma_A1}):
\blk
\begin{equation}
\bar{\Gamma}_{A} = \Gamma_A - E_{\rm cont}(0) ,
\end{equation}
which ensures the bottom of the lower Andreev branch corresponds to $E=0$.
With these modifications, the WKB integrals
(which are central to energy eigenvalue calculations) become
\begin{align}
\sigma(E) = \int_{-\phi_c}^{\phi_c} d\phi \sqrt{  \frac{E - E_\text{cont}(\phi) - \bar \Gamma_A + E_A(\phi)}{4 E_C} }
\label{eq:sigma_E} ,
\end{align}
\begin{align}
\tau(E) = \int_{\phi_c}^{2\pi-\phi_c} d\phi \sqrt{-  \frac{E - E_\text{cont}(\phi) - \bar \Gamma_A + E_A(\phi)}{4 E_C} }
\label{eq:tau_E} ,
\end{align}
\begin{align}
\rho(E) = \int_{-\pi}^{\pi} d\phi \sqrt{-  \frac{E - E_\text{cont}(\phi) - \bar \Gamma_A - E_A(\phi)}{4 E_C} }
\label{eq:rho_E} ,
\end{align}
where the roots $\mp\phi_c$ of the equation
\begin{equation}
E - U_{-}(\phi_c) - \bar \Gamma_A = 0
\label{eq:phi_c} ,
\end{equation}
set the classically available region in the lower branch potential.
Here, the integral $\tau(E)$ corresponds to the forbidden region (under the barrier)
of $U_{-}(\phi)$, while $\rho(E)$ corresponds to moving under the upper branch, $U_{+}(\phi)$,
which is always forbidden. Contributions of the latter are very small \cite{vakhtel_quantum_2023}
and will be neglected below.

\subsubsection{$\sigma(E)$ WKB integral}
In what follows, we will estimate the WKB integrals for $T \lesssim 1$
\red
in
\blk
a transmon-like regime.
Keeping only the leading harmonic in $U_{-}(\phi)$
\red
$\sigma(E)$ simplifies to:
\blk
\begin{equation}
\sigma(E) \simeq
\red
\sqrt{\frac{E_{J,\rm eff}}{4 E_C}}\,
\blk
\int_{-\phi_c}^{\phi_c} d\phi \sqrt{\tilde E + \cos\phi -1}
\label{eq:sigma_E_approx} ,
\end{equation}
where
\red
$\tilde E \equiv \frac{E}{E_{J,\rm eff}}$ and $E_{J,\rm eff}$
\blk
is given by Eq.~(\ref{eq:E_J_eff}).
Expanding $\cos\phi$ to
\red
fourth
\blk
order and integrating over $\phi$
\red
we obtain:
\blk
\begin{eqnarray}
&&\sigma(E) \simeq
\red
\sqrt{\frac{E_{J,\rm eff}}{96 E_C}}\,
\blk
I(u_{-},u_{+}), \ \
u_{\mp} = 6\left[1 \mp \sqrt{1-\frac{2}{3}\tilde E} \right] \quad
\nonumber\\
&& I(u_{-},u_{+}) \equiv \frac{2}{3} \sqrt{u_{+}}\, \left[(u_{+} + u_{-}) E\left(\frac{u_{-}}{u_{+}}\right) \right.
\nonumber \\
&& \qquad\qquad\qquad\qquad \left. { } - (u_{+} - u_{-}) K\left(\frac{u_{-}}{u_{+}}\right) \right]
\label{eq:sigma_E_quartic} ,
\end{eqnarray}
where $K(x)$, $E(x)$ are the complete elliptic integrals. Expanding for small $\tilde E$
\red
we obtain
\blk
\begin{equation}
\sigma(E) \simeq \frac{\pi E}{\omega_{\rm p,eff}} \frac{\left[15+ (1- u_{-}/u_{+})^2\right]}{16 \sqrt{1- \tilde E/6}}
\label{eq:sigma_E_quartic1} ,
\end{equation}
which constitutes a non-linear dependence on $E$.
Here,
\red
\begin{equation}
\omega_{p,\rm eff} = \sqrt{8 E_C E_{J,\rm eff}}
\label{eq:omega_p}
\end{equation}
\blk
is the transmon frequency.
It should be noted, that by expanding $\sigma(E)$, Eq.~(\ref{eq:sigma_E_quartic1}),
to second order and solving the equation
$\sigma(E_n) = \pi (n+ \frac{1}{2})$  for $E_n$ one obtains:
\begin{equation}
E_n =
\red
\omega_{p,\rm eff}
\blk
\left(n+ \frac{1}{2}\right) - \frac{E_C}{12}\,(6 n^2 + 6 n + \frac{3}{2})
\label{eq:E_n_transmon} ,
\end{equation}
which amounts to the transmon anharmonicity: $\alpha_{\rm transmon} = - E_C$.
If not in a transmon limit, the full non-linearity of Eq.~(\ref{eq:sigma_E_quartic})
will leave $\sigma(E_n) \neq \pi (n+ \frac{1}{2})$. %

\subsubsection{$\tau(E)$ WKB integral}
The derivation of the $\tau(E)$ WKB integral \cite{averin_coulomb_1999,vakhtel_quantum_2023}
is relevant in the same regime
of $T \approx 1$.
With the definition
\begin{align}
u(\phi,T) & \equiv \sqrt{1 - T \sin^2\frac{\phi}{2}},
\end{align}
Eq.~(\ref{eq:tau_E}) reads
\begin{equation}
\tau(E)
\approx \sqrt{\frac{\Gamma_A}{E_C}}\,
\int_{\phi_c}^{\pi} d\phi  \sqrt{1  - \frac{E}{\Gamma_A} - u(\phi,\bar T) }
\label{eq:tau_E_approx} ,
\end{equation}
where transparency is renormalized as
\begin{equation}
\bar T \simeq  T\,\left( 1 - \frac{\Gamma_A}{\Delta} \right)
\label{eq:T_bar}
\end{equation}
with the perturbative $E_{\rm cont}(\phi)$ of Eq.~(\ref{eq:E_Cont_pert});
here we neglected
\red
$E_{\rm cont}(0) \ll E_{\rm cont}(\pi)$,
\blk
for $\epsilon_g \ll \Gamma$.
The $\tau_n$-integral
(below $\sigma_n \equiv \sigma(E_n^{(0)})$, $\tau_n \equiv \tau(E_n^{(0)})$)
reads \cite{averin_coulomb_1999,vakhtel_quantum_2023}:
\begin{align}
e^{-\tau_n} & \approx \frac{\sqrt{2\pi}}{n!}
\left( \frac{\bar b^2}{4} \sqrt{\frac{2 \bar T \Gamma_A}{E_C}} \right)^{n + \frac{1}{2} }
e^{-\bar a \sqrt{\frac{2 \bar T \Gamma_A}{E_C}}}
\label{eq:tau_n} ,
\end{align}
where
\begin{align}
\bar a &= \frac{\sqrt{8} \bar r}{(1+ \bar r)\sqrt{1-\bar r}}
\left[ 2\Pi(\bar \mu(0),1,\bar k) - F(\bar \mu(0),\bar k)\right]
\nonumber\\
\bar b &= \lim_{\psi \to 0} \psi e^{ \frac{\sqrt 2 \bar r}{(1 + \bar r)\sqrt{1- \bar r}}
\left[ 2\Pi(\bar \mu(\psi),\frac{1}{\bar k},\bar k) - (1 - \bar r)F(\bar \mu(0),\bar k)\right] }
\nonumber
\end{align}
are the coefficients previously derived in Ref.~\cite{averin_coulomb_1999},
with $\Pi(\varphi,n^2,k)$ and $F(\varphi,k)$ denoting elliptic integrals of
the third and first kind with arguments as given in the notation of
\red
Ref.~\cite{Gradshteyn2014},
\blk
along with the re-scaled ($T \to \bar T$) definitions
\begin{align}
\bar \mu(\phi) & = \arcsin\sqrt\frac{u(\phi,\bar T)-\bar r}{u(\phi,\bar T)+ \bar r},
\nonumber\\
\bar k &= \sqrt\frac{1- \bar r}{1+\bar r},
\nonumber\\
\bar r &= \sqrt{1 - \bar T}.
\end{align}

\subsection{Modification of matching conditions}
For the purposes of charge dispersion (oscillation amplitude of the energy levels with $n_g$)
we will neglect the new charge offsets, $\delta n_g$, $n_z$, introduced in $\hat{H}_{\rm even}$,
and in Eqs.~(\ref{eq:dn_g}), (\ref{eq:n_z})
(these are, however, important for the shifts of energy oscillations, see Sect.~\ref{Sec:charge-dispersion-shift}).

In the region of the minimal gap (at $\phi \approx \pi$) the WKB is not applicable.
Proceeding
\red
similarly
\blk
 to Ref.~\cite{vakhtel_quantum_2023},
\blue
we consider
\blk
the exact solutions
of the Schr\"{o}dinger equation (linearized in this region). %
This amounts to the
\grn
following
\blk
equation:
\begin{equation}
-4 E_C \Psi'' + \tilde{V}_{\pi} \Psi + \bar{\bar{\Gamma}}_A \Psi = 0
\label{eq:linearized_Schrodinger} ,
\end{equation}
where $E_C$ absorbs the capacitance change, Eq.~(\ref{eq:deltaC}),
$\bar{\bar{\Gamma}}_A$, Eq.~(\ref{eq:tildeGamma_A1}), absorbs the continuum contribution at $E_{\rm cont}(\pi)$,
and
\begin{equation}
\tilde{V}_{\pi} = -\tilde{\epsilon}_g\, \tau_x - \frac{1}{2} \tilde{\Gamma}\, (\phi - \pi) \tau_z
- \delta \tilde{\Gamma}\, \tau_y
\label{eq:V_pi} .
\end{equation}
\red
In writing
\blk
Eq.~(\ref{eq:linearized_Schrodinger})
\red
we have also performed
\blk
a $\pi/2$-rotation of the basis
around $\eta_y$ in $\hat{H}_{\rm even}$,  %
\red
such that
\blk
$\eta_x \to -\tau_z$, $\eta_y \to \tau_y$, and $\eta_z \to \tau_x$.

Proceeding similarly to Ref.~\cite{vakhtel_quantum_2023},
\red
we solve
\blk
Eq.~(\ref{eq:linearized_Schrodinger}) with the ansatz
$\Psi = \Psi_{\pi}\, e^{\sigma \bar{\bar k} (\phi - \pi)}$
with $\sigma = \pm 1$ and $\bar{\bar k} = \sqrt{\bar{\bar{\Gamma}}_A/4E_C}$,
and
\red
arrive at
\blk
the (Weber-like) system of equations
for the components of
$\Psi_{\pi} \equiv (u,d)^T$:
\begin{eqnarray}
&&\sigma u' + \sqrt{\bar{\bar \lambda} } d + \frac{1}{2} x u =0
\label{eq:Weber-u}
\\
&&\sigma d' + \sqrt{\bar{\bar \lambda} } u - \frac{1}{2} x d =0
\label{eq:Weber-d} .
\end{eqnarray}
Here, $x = \sqrt{ \tilde{\Gamma}/8 E_C \bar{\bar k} }\, (\phi - \pi)$ is the
\red
rescaled
\blk
position,
and
\begin{equation}
\bar{\bar \lambda} = |\tilde r|^2\, \tilde{\Gamma}/{8 E_C \bar{\bar k} }, \ \ \
\tilde r = \frac{-\tilde{\epsilon}_g + i \delta{\tilde{\Gamma}} }{\tilde \Gamma}
\label{eq:lambda-derivation} .
\end{equation}
This constitutes the derivation of Eq.~(\ref{eq:lambda-parameter}) of the main text.
\red
Equations
\blk
~(\ref{eq:Weber-u}), (\ref{eq:Weber-d}) amount to
\grn
the
\blk
Weber equations for $u(x)$
\blue
and
\blk
$d(x)$
and coincide in their form with that of Ref.~\cite{vakhtel_quantum_2023}.
Asymptotics of their exact solutions are matched with the corresponding WKB solutions
(at $T\approx 1$, i.e. for $\epsilon_g, \delta\Gamma \ll \Gamma$)
with analogous
\blue
rescaling
\blk
of the parameters.
This leads to the following equation
connecting the WKB integrals \cite{vakhtel_quantum_2023}
(assuming $e^{2\rho},\, e^{2\tau} \gg 1$):
\begin{align}
\cos\sigma
\approx e^{-\rho} e^{-\tau} \cos(4\pi n_g) + w(\bar{\bar \lambda}) e^{-\tau} \cos(2\pi n_g + \tilde\delta)
\label{eq:Vakhtel-B781} .
\end{align}
Here, $\tilde\delta$, $w(\bar{\bar \lambda})$, are functions of $\epsilon_g$, $\Gamma$, $\delta\Gamma$, $\Delta$,
and are given by
Eqs.~(\ref{eq:WKB-delta E_n}) and (\ref{eq:w-amplitude-lower-branch})
of the main text,
respectively.
Below it is shown that $\tilde\delta$ is the shift of energy oscillations as a function of $n_g$
in the limit
of small $\epsilon_g$, $\delta\Gamma$.
The quantity $w(\bar{\bar \lambda})$ has the interpretation of
\grn
the
\blk
\blue
probability
\blk
amplitude of remaining on the lower
Andreev branch \cite{averin_coulomb_1999,vakhtel_quantum_2023} while the system is tunneling
under the barrier around $\phi = \pi$;
$w(\bar{\bar \lambda})$ will be useful in the interpretation of the shift of energy oscillations
for large $\epsilon_g$, $\delta\Gamma$, where approximate
solutions of the eigenvalue equation with $\hat{H}_{\rm even}$
are given in Sect.~\ref{Sec:charge-dispersion-shift}.

\subsection{Solving Eq.~(\ref{eq:Vakhtel-B781})}

\grn
In the following
\blk
we obtain an approximate solution of
the transcendental equation Eq.~(\ref{eq:Vakhtel-B781}),
valid for $T \approx 1$,
where the term $\propto e^{-\rho} \ll 1$ will be neglected.
In the transmon limit where $e^{-\tau_n}$ is exponentially suppressed,
Eq.~(\ref{eq:tau_n}),
\red
we obtain
\blk
$\cos\sigma_n \simeq 0$, and
\begin{equation}
\sigma_n = \pi \left(n + \frac{1}{2} \right)
\end{equation}
from which one can infer the standard transmon anharmoniciy
via Eq.~(\ref{eq:E_n_transmon}).
In a non-transmon regime, based on Eq.~(\ref{eq:sigma_E_quartic}),
we will assume $\sigma_n \neq  \pi (n + \frac{1}{2})$.

\red
We use
\blk
the Newton-Raphson method to solve Eq.~\eqref{eq:Vakhtel-B781} for
$E_n$,
\red
starting
\blk
with the initial guess $E_n^{(0)}$ given by
the harmonic approximation
\begin{align}
E_n^{(0)} &=
\red
\omega_{p, \rm eff}
\blk
\left( n + \frac{1}{2} \right)
\end{align}
At $j$-th iteration, the approximate energy eigenvalue $E_n^{(j)}$ is given by
\begin{align}
E_n^{(j+1)} = E_n^{(j)} - \frac{f\left(E_n^{(j)}\right)}{f'\left(E_n^{(j)}\right)} ,
\end{align}
where
\begin{align}
f(E) = \cos\sigma(E) - w(\bar{\bar \lambda}) e^{-\tau(E)} \cos(2\pi n_g + \tilde \delta) .
\label{eq:Newton-Raphson-f(E)}
\end{align}
By neglecting the WKB exponent in the denominator,
\red
we obtain:
\blk
\begin{align}
E_n^{(j+1)} & \approx  E_n^{(j)}
\label{eq:En-Newton-Raphson}\\
& { } + \frac{ \cos[\sigma(E_n^{(j)})]
- w(\bar{\bar \lambda}) e^{-\tau(E_n^{(j)})} \cos(2\pi n_g + \tilde \delta)  }{ \sin[\sigma(E_n^{(j)})]\, \sigma'(E_n^{(j)}) }.
\nonumber
\end{align}
We define the anharmonicity at $j$-th iteration as
\begin{align}
\alpha^{(j)} = (E_2^{(j)} - E_1^{(j)}) - (E_1^{(j)} - E_0^{(j)})
\label{eq:anharmonicity-WKB} .
\end{align}
In a deep transmon-like regime ($E_C \ll \bar T \Gamma_A$) for which the $n_g$-dependent oscillations
are exponentially suppressed,
\blue
and for low transparency
\blk
the above numerical iteration procedure with $j=2$ leads
to
\red
a good approximation for the
\blk
anharmonicity, $\approx -E_C$, where the
\red
initial guess
\blk
 for $\cos [\sigma(E_n^{(0)})]$ is via
Eq.~(\ref{eq:sigma_E_quartic}).
\red
We use
\blk
Eq.~(\ref{eq:En-Newton-Raphson})
\red
with two iterations
\blk
to calculate numerically the anharmonicity in the non-transmon regime
of Fig.~\ref{fig:anharmonicity-high-T-ng-01} and Fig.~\ref{fig:anharmonicity-high-T-ng-06}
of the main text,
which reproduces the overall behavior for small $\epsilon_g \ll \Gamma$,
for $\alpha(n_g=0,\epsilon_g)$ vs. $\alpha(n_g=0.5,\epsilon_g)$, as well as
the intercept at $\alpha(\epsilon_g=0)$.

Note, that in a transmon-like regime,
where $\cos \sigma_n \simeq 0$,
Eq.~(\ref{eq:En-Newton-Raphson})
reproduces the energy eigenvalue corrections,
$\delta E_n$, as of Refs.~\cite{averin_coulomb_1999,vakhtel_quantum_2023}
\begin{align}
\delta E_n &\approx \delta E_n^{(1)} + \delta E_n^{(2)}
\nonumber\\
\delta E_n^{(1)} & = (-1)^{n+1} e^{-\tau(E_n^{0})}
\frac{w(\bar{\bar \lambda})  \cos(2\pi n_g + \tilde \delta) }{  \sigma'(E_n^{0}) }
\label{eq:N-R-1st-iteration}\\
\delta E_n^{(2)} &\approx
\left[\frac{ w(\bar{\bar \lambda}) e^{-\tau(E_n^{0})} }{\sigma'(E_n^{0})} \cos(2\pi n_g + \tilde \delta)\right]^2 \tau'(E_n^{0}),
\label{eq:N-R-2nd-iteration} ,
\end{align}
where $\delta E_n^{(1)}$ corresponds to Eq.~(\ref{eq:WKB-delta E_n}) of the main text.

\bibliography{zotero,extra}

\end{document}